\documentclass[aip,cha,reprint]{revtex4-2}

\usepackage{newtxtext}
\usepackage{amsmath}
\usepackage{amssymb}
\usepackage[T1]{fontenc}
\usepackage{graphicx}
\usepackage{dcolumn}
\usepackage{mathtools}
\usepackage{relsize}
\usepackage{subcaption}
\usepackage{braket}
\usepackage{booktabs}
\usepackage{multirow}
\usepackage{xcolor}
\usepackage{etoolbox}
\usepackage{bm} 

\newcommand{\evpatm}{$\rm{e\!V}/\text{atom}$}
\newcommand{\evpang}{$\rm{e\!V}/\text{\AA}$}
\newcommand{\fs}{\footnotesize}


\begin{document}

\title{Composable and adaptive design of machine learning interatomic potentials guided by Fisher-information analysis}

\author{Weishi Wang}
\email{weishi.wang.gr@dartmouth.edu}
\affiliation{Lawrence Livermore National Laboratory, Livermore, CA 94550, USA}
\affiliation{Department of Physics and Astronomy, Dartmouth College, Hanover, NH 03755, USA}

\author{Mark K. Transtrum}
\affiliation{Department of Physics and Astronomy, Brigham Young University, Provo, UT 84602, USA \looseness=-1}

\author{Vincenzo Lordi}
\affiliation{Lawrence Livermore National Laboratory, Livermore, CA 94550, USA}

\author{Vasily V. Bulatov}
\affiliation{Lawrence Livermore National Laboratory, Livermore, CA 94550, USA}

\author{Amit Samanta}
\affiliation{Lawrence Livermore National Laboratory, Livermore, CA 94550, USA}

\begin{abstract}
An adaptive physics-inspired model design strategy for machine-learning interatomic potentials (MLIPs) is proposed. This strategy relies on iterative reconfigurations of composite models from single-term models, followed by a unified training procedure. A model evaluation method based on the Fisher information matrix (FIM) and multiple-property error metrics is also proposed to guide the model reconfiguration and hyperparameter optimization. By combining the reconfiguration and the evaluation subroutines, we provide an adaptive MLIP design strategy that balances flexibility and extensibility. In a case study of designing models against a structurally diverse niobium dataset, we managed to obtain an optimal model configuration with 75 parameters generated by our framework that achieved a force RMSE of 0.172~{\evpang} and an energy RMSE of 0.013~{\evpatm}.
\end{abstract}

\maketitle

\section{Introduction}
The potential energy surface (PES) of a material plays a crucial role in determining its macroscopic physical properties. Various approaches, from wavefunction-based electronic structure methods such as coupled cluster singles and doubles (CCSD)~\cite{helgaker2013molecular} to density functional theory (DFT)~\cite{martin2020electronic}, have been proposed to probe the PES, each with different levels of approximation and computational efficiency. Although DFT methods are less computationally costly than multi-configurational wavefunction-based methods, they still become expensive for systems with more than a few hundred atoms. In addition to these {\it ab initio} approaches, machine learning interatomic potentials (MLIP)~\cite{bartok2010gaussian, behler2016perspective} have emerged as a new class of methods that strike a balance between computational efficiency and accuracy, enabling the study of PESs in multi-element systems with larger sizes.

To obtain a general expression of the interatomic potential for a target system, an MLIP method defines a parameterized model $M$ that approaches a ground-truth (GT) PES as it is being trained against a set of GT energies and/or atomic forces. Due to this data-driven approach, it is crucial that the dataset provides enough information to model the ground truth PES, i.e., the dataset is sufficiently diverse and uniformly dispersed over relevant portions of the configuration or latent space.\cite{loukas2024generalizingdiversedistributionuniformity,sun2023exploring} Additionally, the proposed MLIP model $M$ must be expressive enough to generate a functional form close to the ground truth with converged model parameters. Early MLIPs used relatively simple artificial neural network (NN) architectures~\cite{behler2007generalized, behler2011neural, behler2014representing, zhang2020dp}, but more recent MLIPs based on graph neural networks (GNNs), with specific architectures to incorporate intrinsic symmetries and higher-order many-body correlations, have the complexity required to learn PESs of a wide variety of materials~\cite{xie2018crystal, chen2019graph, yang2023graph, batzner20223}. A key conclusion from this evolution in the complexity of MLIPs in the past years is that generic machine learning architectures may adequately capture relevant many-body correlations needed to model energies obtained from quantum mechanical calculations. This raises a natural question: Is it possible to design tailored physics-inspired analytic models that have more flexible functional forms than the traditional embedded atom method (EAM)~\cite{daw1984embedded} and Stillinger--Weber~\cite{vink2001fitting, zhou2013stillinger} potentials, yet require fewer parameters than GNNs to include higher-order correlations? One possible approach is an adaptive design strategy that combines physics-inspired models and necessary nonlinear and collective interactions between atoms. This approach can lead to models with an optimal (smaller) number of trainable parameters, simplified and interpretable architecture, and improved performance~\cite{lin2022development}.

The complex architectures of MLIPs proposed in recent years also introduce additional challenges regarding trainability. On the one hand, to ensure the model's functional space is large enough to include the desired GT subspace, thousands, and often millions, of parameters are required to construct the model~\cite{gasteiger2021gemnet, batzner20223, batatia2022mace}. This leads to a significant increase in training cost and complexity of finding the global training minimum, as well as significant computational resources needed to generate the large GT datasets via DFT or higher fidelity calculations~\cite{fung2021benchmarking, kocer2022neural}. Recent analysis of loss function landscapes of deep learning models also showed that the majority of eigenvalues of the Hessian of the loss function are close to zero~\cite{gur2018gradient, sagun2016eigenvalues, pennington2017geometry}, and the relevance of such degrees of freedom to the performance of an MLIP is difficult to assess. On the other hand, we know from the Cram\'er--Rao bound~\cite{kurniawan2022bayesian} that the variance of predictions obtained from an MLIP is bounded by the inverse of the Fisher information matrix (FIM). Therefore, it is natural to seek a possible strategy that uses the FIM to minimize systematic model bias.

Given these limitations of deep neural networks~\cite{poggio2020theoretical,mao2024training}, one can also ask: Are large and complex machine-learning architectures necessary to generate accurate data-driven interatomic potentials? Admittedly, the conventional empirical models like Lennard--Jones potential~\cite{hansen1969phase} and EAM generally perform worse than the much larger NN models. Nevertheless, a number of physics-inspired MLIP models have also been proposed~\cite{baskes1992modified, bartok2010gaussian, thompson2015spectral, shapeev2016moment, drautz2019atomic} which share a similar philosophy with those simple conventional models yet provide comparable performance against the NN models~\cite{zuo2020performance}.

\begin{figure}[htp]
    \centering
    \includegraphics[width=0.45\textwidth]{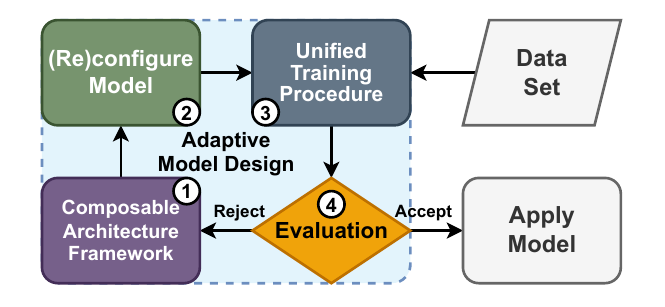}
    \caption{An adaptive MLIP model design procedure is shown in the above diagram. Step 1: Choose a composable architecture framework that supports iterative combinations of submodel architectures and basis functions based on specified configurations. Step 2: Assemble an initial model configuration. Step 3: Train the model configuration based on a unified procedure. Step 4: Evaluate the performance of the model configuration based on both the training RMSEs and the Fisher information matrix. Based on the evaluation result, if the model configuration is accepted, it is then applied to test sets. Otherwise, by switching on and off the components in the framework (Step 1), a reconfigured model is generated (Step 2), trained (Step 3), and evaluated again (Step 4). So on and so forth, a final model configuration reaches the balance between performance and efficiency.}
    \label{fig:adaptivedesign}
\end{figure}

To mitigate some of the issues related to MLIPs and inspired by the success of physics-inspired models, this work explores a model design and training framework that provides an adaptive optimization of the model configuration during the training stage. By utilizing a composable model architecture framework guided by a Fisher-information-based evaluation strategy, we iteratively readjust the model configuration (e.g., hyperparameters, structures) and systematically improve its accuracy using a unified training procedure. A schematic of our approach is shown in FIG.~\ref{fig:adaptivedesign}.

The paper is organized as follows. In Sec.~\ref{sec:stm}, we introduce a unified model composition framework that allows generating composite model configurations from a set of ``single-term'' models. Utilizing these basic components within the framework, we propose several ``multi-term'' models, based on the architectures with emphasis on co-optimizing the model FIM eigenspectrum and training accuracy in Sec.~\ref{sec:dtm}. In Sec.~\ref{sec:results}, against a niobium training dataset, we demonstrate the effectiveness of a complementary model evaluation strategy based on the FIM and four property-oriented error metrics for model performance estimation. Furthermore, we show how this strategy can guide the model reconfiguration through hyperparameter tuning. Finally, we conclude our paper in Section~\ref{sec:co} with discussions about the characteristics of specific model architectures proposed in this paper, as well as future directions to follow our research.

\section{Single-term and composable interatomic potentials}\label{sec:stm}
The essence of a composable model design relies on the balance between flexibility and extensibility. We divide the general procedure of constructing an MLIP model into two stages. The first stage is building an expressive basis set from physics-inspired feature functions (i.e., descriptors). The second stage is instantiating the submodel architectures with these functions as reusable components. 

Formally, we propose a (finite) many-body cluster \textbf{basis set} $\mathfrak{F}(\bm{\theta}_{\mathfrak{F}})$ as the basic units for generic MLIP model composition. Specifically, each cluster basis function $\psi_{n{\rm b},\nu} \in \mathfrak{F}$ represents the $\nu$th fragment of an $n$-body correlation within a local atomistic environment. For the formalism of $\psi_{n{\rm b},\nu}$, we refer the reader to Appendix~\ref{app:lcb}. Correspondingly, we define an MLIP \textbf{architecture} $\hat{M}(\bm{\theta}_M)$ as a parameterized operator acting on $\mathfrak{F}(\bm{\theta}_{\mathfrak{F}})$ to produce an interatomic potential \textbf{model} configuration:
\begin{equation}
            \hat{M}[\mathfrak{F}] \equiv \hat{M} ( \bm{\theta_{M}},\, \mathfrak{F} ( \bm{\theta}_{\mathfrak{F}} )).
\end{equation}
As a scalar function, $\hat{M}[\mathfrak{F}]$ takes in the encoded positions $\bm{\Lambda_i}$ of the atoms around a center atom indexed by $i$ (e.g., an atomic neighbor list) and outputs the averaged interatomic potential with respect to that atom:
\begin{equation}\label{eq:model_config}
E_{M[\mathfrak{F}],i}\!\left(\bm{\theta_{M[\mathfrak{F}]}}\right) \!=\! \hat{M}[\mathfrak{F}]\!\left(\bm{\Lambda_i}\right)\;\mathrm{with}\hspace{0.4em}\bm{\theta_{M[\mathfrak{F}]}} \!=\! \bm{\theta_{M}} \cup \bm{\theta}_{\mathfrak{F}},
\end{equation}
where $\bm{\theta_{M[\mathfrak{F}]}}$ denotes all the unique parameters (distinguished by their associated symbols) $E_{M[\mathfrak{F}]}$ inherits from the applied architecture and basis set. Finally, summing $E_{M[\mathfrak{F}],\,i}$ over all the center atoms in a target (unit cell) structure returns the atomistic approximation of its total energy
\begin{equation}\label{eq:pes1}
        E_{\rm tot}\left(\bm{r_1}^{\braket{e_1}},\dots,\bm{r_{N_{\rm a}}}^{\braket{e_{N_{\rm a}}}}\right) = \sum_{i=1}^{N_{\rm a}} E_{M[\mathfrak{F}],i}\!\left(\bm{\theta_{M[\mathfrak{F}]}}\right),
\end{equation}
where $e_i$ is the atomic symbol of the $i$th atom, with the total particle number being $N_{\rm a}$. To avoid confusion, superscripts enclosed by $\braket{\,}$ throughout this paper are considered labels rather than exponents.

\subsection{Linear models}\label{sec:lgp}
One of the simplest and most direct machine-learning model architectures is taking the linear combinations of basis functions from a basis set. This approach has been used to construct models such as moment tensor potentials (MTP)~\cite{shapeev2016moment, podryabinkin2017active} and atomic cluster expansions (ACE)~\cite{drautz2019atomic} with specific local descriptors as basis functions. In the many-body cluster formalism, we can construct an $n$th-order basis set 
\begin{equation}
    \mathfrak{F}_{n{\rm b}}\bigl(\bm{\theta_{\mathfrak{F}_{n{\rm b}}}}\bigr) \equiv \left\{ \psi_{n{\rm b},\nu}  \,|\, \nu=1,\dots,N_{\rm 2b} \right\},
\end{equation}
where each $\psi_{n{\rm b},\nu}(\mathcal{N}_i;\, \bm{\theta_{\psi_{n{\rm b},\nu}}})$ is an $n$-body cluster basis function, parameterized by $\bm{\theta_{\psi_{n{\rm b},\nu}}} \subseteq \bm{\theta_{\mathfrak{F}_{n{\rm b}}}}$, that takes an atomic neighbor list $\mathcal{N}_i$ as the input. Formally, we define $\mathcal{N}_i$ by:
\begin{equation}\label{eq:neighbor_list}
    \mathcal{N}_i \equiv \bigl\{\bm{r_{ij}}\,\big|\,j \!\in\! \{1,\dots,N_{\rm a}\,|\,j\!\neq\!i; \;\lVert\bm{r_{ij}}\rVert_2 \!\leq\! r_{\rm c}\}\bigr\}, 
\end{equation}
where $r_{\rm c}$ is the cutoff radius, and $\bm{r_{ij}} \equiv \bm{r_i}^{\braket{e_i}} \!-\! \bm{r_j}^{\braket{e_j}}$ is the interatomic displacement. We note that $\bm{r_i}^{\braket{e_i}}$ and $\bm{r_{ij}}$ are both vectors, as their indices mark the associated atoms, not the component along a certain dimension. Consistently, we apply the bold font to any vectorized or collection-like variables throughout this paper, as long as their subscripts (or superscripts) do not index a scalar component.

The linear combination of $\psi_{n{\rm b},\nu}$ with the same $n$ forms a model that directly accounts for the $n$-body interactions. More concretely, in the case of $n\!=\!2$, we define the pair-interaction potential by
\begin{equation}\label{eq:l2b}
    \begin{aligned}
        &E_{S_{\, {\rm l}2}[\mathfrak{F}_{\rm 2b}],i}\!\left(\bm{\theta_{S_{\, {\rm l}2}[\mathfrak{F}_{\rm 2b}]}}\right) = \sum_{\nu=1}^{N_{\rm 2b}}c_{\nu}\,\psi_{2{\rm b},\nu}\!\left(\mathcal{N}_i;\,\bm{\theta_{\psi_{2{\rm b},\nu}}}\right) + \tau^{\braket{e_i}}\\
        &\mathrm{with}\hspace{0.5em}\bm{\theta_{S_{\, {\rm l}2}[\mathfrak{F}_{\rm 2b}]}} = \bm{c} \cup \bm{\tau}^{\braket{\mathfrak{A}}} \cup \bm{\theta_{\mathfrak{F}_{\rm 2b}}},
    \end{aligned}
\end{equation}
where $\hat{S}_{{\rm l}2}$ is the underlying model architecture, parameterized by the linear coefficients $\bm{c} = \{c_\nu\,|\,\nu\}$ and a shifting parameter set $\bm{\tau}^{\braket{\mathfrak{A}}} \equiv \{\tau^{\braket{e}}\,|\,e \in \mathfrak{A}\}$, where $\mathfrak{A}$ are the atomic symbols associated with the target system. Since $\psi_{n{\rm b},\nu}$ encodes both the symmetry and the chemical information of the local environment (see Appendix \ref{app:lcb}), Equation~(\ref{eq:l2b}) can be easily extended to a linear $n$-body-interaction model $\hat{S}_{\, {\rm l}n}$ with $n > 2$. Combining the two-body and three-body linear potentials with the EAM potentials, the generalized EAM (GEAM) models have been shown to perform significantly better~\cite{sharma2023development, sun2023exploring} than the original EAM formalism. Moreover, the strategy of combining potentials that capture two-body and three-body interactions to construct composite potential models (e.g., the Stillinger--Weber potential~\cite{vink2001fitting, zhou2013stillinger}) has also been shown to be effective in MLIP design~\cite{drautz2019atomic}.

We categorize the models that can be used as basic components (submodels) to construct composite models as \textbf{single-term} potential models. In contrast, we define the \textbf{multi-term} interatomic potential models as any $\hat{M}[\mathfrak{F}]$ with the identities:
\begin{equation}\label{eq:pes3b}
    \begin{aligned}
        &\hat{M} [ \mathfrak{F} ] \!=\! \hat{D} \bigl[ \hat{M}_{\rm L}\left[ \mathfrak{F}_{\rm L}\right]\!,\hat{M}_{\rm R}\left[\mathfrak{F}_{\rm R}\right]\bigr]
        \!=\! \left(\hat{M}_{\rm L}\left[ \mathfrak{F}_{\rm L}\right]\right) \!\hat{D}\! \left(\hat{M}_{\rm R}\left[ \mathfrak{F}_{\rm R}\right] \right)\\        &\mathrm{with}\hspace{0.5em}\mathfrak{F} = \mathfrak{F}_{\rm L} \cup \mathfrak{F}_{\rm R},\;\;\bm{\theta_{M\left[ \mathfrak{F}\right]}} = \bm{\theta_{M_{\rm L}\left[ \mathfrak{F}_{\rm L}\right]}} \cup \bm{\theta_{M_{\rm R}\left[ \mathfrak{F}_{\rm R}\right]}},
    \end{aligned}
\end{equation}
where $\hat{D}$ is a binary (dual-model) operator that combines the sub-models of $\hat{M} [\mathfrak{F}]$: $\hat{M}_{\rm L}\left[\mathfrak{F}_{\rm L}\right]$ and $\hat{M}_{\rm R}\left[\mathfrak{F}_{\rm R}\right]$. Recursively, $M_{\rm L}$ and $M_{\rm R}$ may be further decomposed until the subsequent submodels are single-term models $\hat{S}_{\Box}[\mathfrak{F}_{S_{\Box}}]$. Through a sequence of applying the binary operators $\{\hat{D}_i\,|\,i\}$, one may construct a \textbf{composable} MLIP model hierarchically from the single-term models. In the case where $\hat{D}_i$ are parameterized by $\bm{\theta_{D_i}}$, those parameters shall be included in the final parameter set of $\hat{M}[\mathfrak{F}]$, i.e., $\bigcup_i\bm{\theta_{D_i}} \subseteq \bm{\theta_{M[\mathfrak{F}]}}$. The linearity of a single-term model architecture can be further categorized based on the relationship between its coefficients and the basis functions it applies to. For instance, $\hat{S}_{\, {\rm l}n}$ is considered a \textbf{linear} architecture, hence its output  $\hat{S}_{\, {\rm l}n}[\mathfrak{F}_{n{\rm b}}]$ is a \textbf{linear single-term model}.

In general, when a numerical model's (vectorized) parameters $\bm{\theta}$ are fitted by linear or nonlinear regression methods, the sensitivity of the model's prediction with respect to $\bm{\theta}$ is quantified by the Fisher information matrix (FIM). The FIM eigenspectrum provides valuable geometric information about the loss (or likelihood) function landscape and how well different directions in the parameter space are constrained by the data. On the one hand, large eigenvalues correspond to the directions that are well-constrained by the data, indicating the model has low uncertainty along these eigenvectors. On the other hand, the small eigenvalues of FIM correspond to poorly constrained parameter directions, i.e., directions along which the loss function is relatively flat and has high uncertainty. For MLIP development, the standard least-squares loss is often used as the negative log-likelihood (NLL) function to quantify discrepancies between GT and model predictions for the total energies and the interatomic forces~\cite{machta2013parameter,kurniawan2024information}. Consider a one-dimensional NLL (up to an irrelevant constant):
\begin{equation}
 {\rm{NLL}}_{\mathcal{M}}\!\left(\bm{\theta}\right) = \sum_{s} w_{s} \left|p_{s} - \mathcal{M}\!\left(x_{s};\,\bm{\theta}\right)  \right|^{2},
 \label{eq:likelihoodregression}
\end{equation}
where $w_{s}$ is the fitting weight for the $s$th GT value $p_{s}$, and $\mathcal{M}\!\left(x_{s};\,\bm{\theta}\right)$ is the model prediction parameterized by $\bm{\theta}$, evaluated at data point $x_s$. The Fisher information corresponding to ${\rm{NLL}}_{\mathcal{M}}\!\left(\bm{\theta}\right)$ is defined as
\begin{equation}\label{eq:fim}
 \mathcal{I}^{\braket{\mathcal{M}}}\!\left(\bm{\theta} \right) \equiv \Big\langle \frac{\partial^{2} {\rm{NLL}}_{\mathcal{M}}\!\left(\bm{\theta}\right)}{\partial\bm{\theta}\partial\bm{\theta}^\mathsf{T}} \Big\rangle_{\xi_s}
\end{equation}
where $\braket{\cdot}_{\xi_s}$ denotes the expectation value with respect to theoretical values of $p_s$, assuming they deviate from the predictions $\mathcal{M}\!\left(x_s;\,\bm{\theta}\right)$ by a Gaussian random variable $\xi_s = p_s - \mathcal{M}\!\left(x_s;\,\bm{\theta}\right)$ with zero mean and variance $w_s^{-1}$. 

For linear models, $\mathcal{M}\!\left(x_{s};\,\bm{\theta}\right)$ generalizes to linear transformations of the descriptors (e.g., basis functions) characterized by linear coefficients $\bm{c} \subseteq \bm{\theta}$. Accordingly, the FIM for least-squares regression~\cite{transtrum2011geometry} with respect to $\bm{c}$ becomes
\begin{equation}
    \mathcal{I}^{\braket{P}}_{ij} = 2 \sum_{s} w_s \frac{\partial \mathcal{M}\!\left(x_{s};\,\bm{\theta}\right)}{\partial c_i} \frac{\partial \mathcal{M}\!\left(x_{s};\,\bm{\theta}\right)}{\partial c_j}.
\end{equation}
Therefore, the FIM for a linear single-term model with respect to its coefficients is directly related to the ability of its basis functions to encode interatomic environments. When basis functions have redundancies or are unable to fully capture higher-order many-body correlations in the distribution of neighbors around an atom, the FIM becomes ill-conditioned and contains many sloppy modes~\cite{parsaeifard2022manifolds}.

\subsection{Nonlinear models}\label{sec:nlp}
Since a trained model is uncertain in the parameters aligned with the small eigenvalues of the FIM, one can reduce systematic bias and uncertainties by tuning the functional form of the model to selectively remove those eigenvalues. To design models with co-optimization of the FIM eigenspectrum, in this sub-section, we explore the possibility of applying nonlinear transformations to basis functions to construct \textbf{nonlinear single-term} models. Similar to the 
 case discussed in Section~\ref{sec:lgp}, consider a one-dimensional model prediction as the nonlinear mapping $g$ from the outputs of a set of local descriptors $\psi_{\nu}(x;\,\bm{\theta_{\nu}})$:
\begin{equation}
    \begin{aligned}
        &\mathcal{M}\!\left(x_{s};\,\bm{\theta}\right) \coloneqq g\bigl(\{\psi_{\nu}(x_s;\,\bm{\theta_{\nu}})\,|\,\nu\};\,\bm{\theta_g}\bigr)\\
        &\mathrm{with}\hspace{0.5em} \bm{\theta} = \left(\cup_{\nu}\bm{\theta_{\nu}}\right) \cup \bm{\theta_g}.
    \end{aligned}
\end{equation}
Then, according to Equation (\ref{eq:fim}), the FIM corresponding to the NLL defined in Equation (\ref{eq:likelihoodregression}) becomes 

\begin{equation}\label{eq:nlst}
 \mathcal{I}^{\braket{\mathcal{M}}}_{ij} \!= 2\!\sum_{s}w_{s}\frac{\partial}{\partial\theta_j}\!\!\left[\bigl(\mathcal{M}\!\left(x_{s};\,\bm{\theta}\right)\!-\!p_{s} \bigr)\frac{\partial \mathcal{M}\!\left(x_{s};\,\bm{\theta}\right)}{\partial \theta_i} \right].
\end{equation}
Equation~(\ref{eq:nlst}) can be generalized to computing the FIM of a single-term model with respect 
 to any type of coefficients or hyperparameters. An important follow-up question is: What analytic form should $g$ take such that it can significantly alter the distribution of FIM eigenvalues? Motivated by the success of the NN potential models in which exponential functions are used to incorporate nonlinearity~\cite{behler2007generalized, behler2011neural, behler2014representing}, we propose an exponentiated pair-cluster potential:
\begin{equation}\label{eq:e2b}
    \begin{aligned}
        &E_{S_{{\rm e}2}[\mathfrak{F}_{\rm 2b}],i} \!\left(\bm{\theta_{S_{{\rm e}2}[\mathfrak{F}_{\rm 2b}]}}\right) \!\equiv \!\hspace{-1.0em}\sum_{\mu=1,\;\nu=1}^{N_\zeta,\;N_{\rm 2b}}\hspace{-0.8em}\!c_{\mu\nu} \!\exp\Bigl(\!-\zeta_{\mu}\psi_{2{\rm b},\nu}\!\left(\mathcal{N}_i;\bm{\theta_{\psi_{2{\rm b},\nu}}}\right)\!\Bigr)\\
    &\mathrm{with}\hspace{0.5em}\bm{\theta_{S_{{\rm e}2}[\mathfrak{F}_{\rm 2b}]}} = 
        \bm{c} \cup 
        \bm{\mu} \cup \bm{\theta_{\mathfrak{F}_{\rm 2b}}},
    \end{aligned}
\end{equation}
where $c_{\mu\nu} \!\in\! \bm{c}$ are linear coefficients same to which are in $\hat{S}_{{\rm l}2}$, except they span an additional dimension along index $\mu$. This additional parameter dimension in $\hat{S}_{{\rm e}2}$ correlates a set of exponent coefficients $\zeta_{\mu}$. They span a latent space where each basis function is mapped to an exponential function. Furthermore, similar to the inclusion of $\tau^{\braket{e_i}}$ in $\hat{S}_{{\rm l}2}$, we can also ``shift'' the output potential of $\hat{S}_{{\rm e}2}$ by a translation map:
\begin{equation}\label{eq:shift}
    E_{\Box,i}\left(\bm{\theta_{\Box}}\right) \mapsto E_{\Box,i}\left(\bm{\theta_{\Box}}\right) + \tau^{\braket{e_i}}.
\end{equation}
Any operator $\hat{O}$ (e.g., a single-term model architecture) whose output function is shifted by a scalar parameter shall be appended with an additional right subscript ``$[+]$'' throughout this paper. For instance, the shifted exponentiated pair-cluster architecture is denoted as $\hat{S}_{{\rm e2}[+]}$.

\begin{figure*}[ht]
    \centering
    \begin{subfigure}[t]{0.435\textwidth}
        \centering
        \includegraphics[width=0.7941\textwidth]{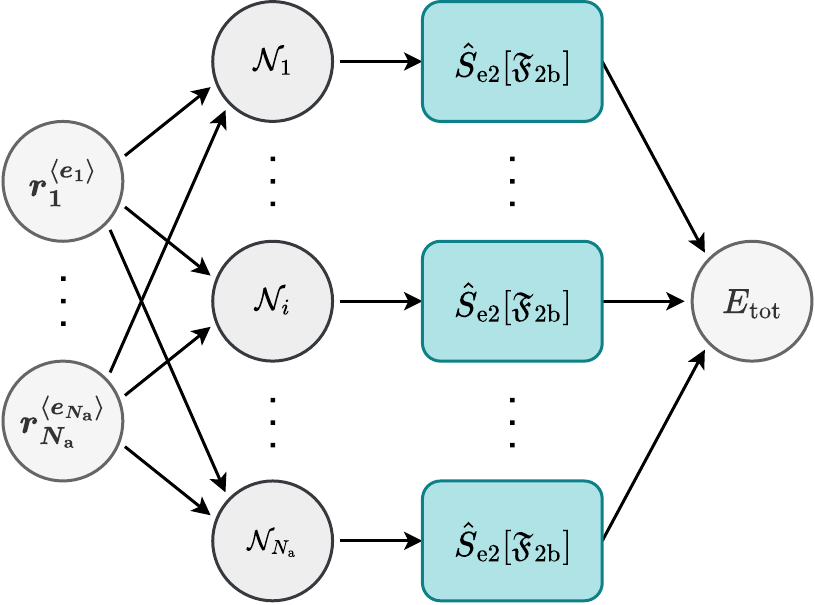}
        \caption{The computation graph of applying $\hat{S}_{{\rm e}2}[\mathfrak{F}_{\rm 2b}]$ to compute the total energy of a target structure. The first two layers transform the atomic coordinates into atomic neighbor lists $\mathcal{N}_i$. Next, each neighbor list goes through $\hat{S}_{{\rm e}2}[\mathfrak{F}_{\rm 2b}]$ in the third layer. Finally, the outputs are summed up.}
        \label{fig:e2b1}
    \end{subfigure}
    \hfill
    \begin{subfigure}[t]{0.45\textwidth}
        \centering
        \includegraphics[width=\textwidth]{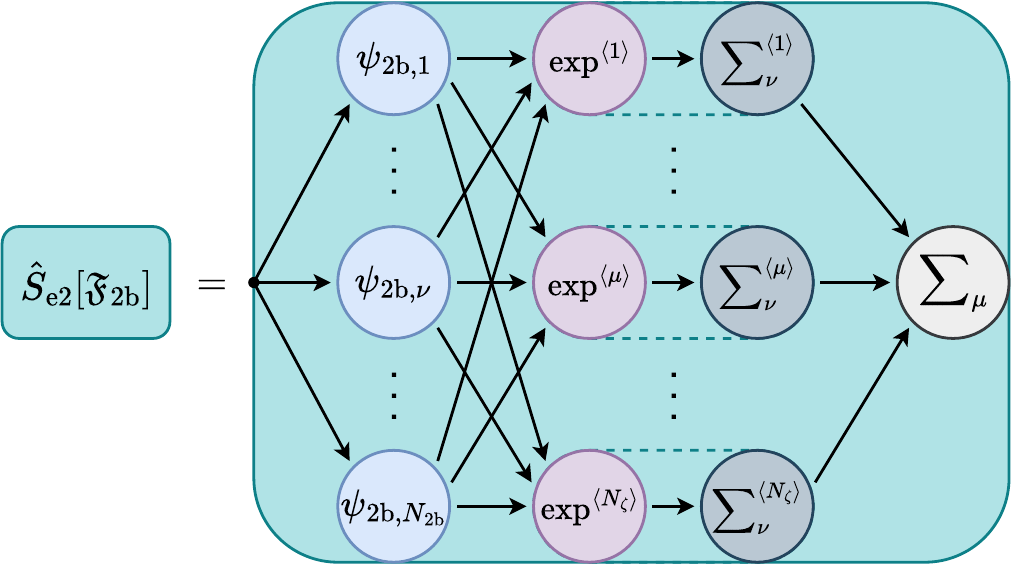}
        \caption{The computation graph of $\hat{S}_{{\rm e}2}[\mathfrak{F}_{\rm 2b}]$. $\psi_{n{\rm b},\nu}$ in the first layer generate two-body interaction fragments $x_{\nu}$ based on the input $\mathcal{N}_i$. Then, the nodes in the second layer are passed through two chained mappings, $\exp^{\braket{\mu}}\!\!:x_{\nu}\!\mapsto\!\exp(\zeta_{\mu}x_{\nu})$ and ${\sum}^{\braket{\mu}}_{\nu}\!\!: y_{\nu}\!\mapsto\!\sum_{\nu}c_{\mu\nu}y_{\nu}$. Finally, the node in the last layer performs a contraction on the outputs from the second layer with respect to index $\mu$.}
        \label{fig:e2b2}
    \end{subfigure}
    \begin{subfigure}[t]{0.435\textwidth}
        \centering
        \includegraphics[width=\textwidth]{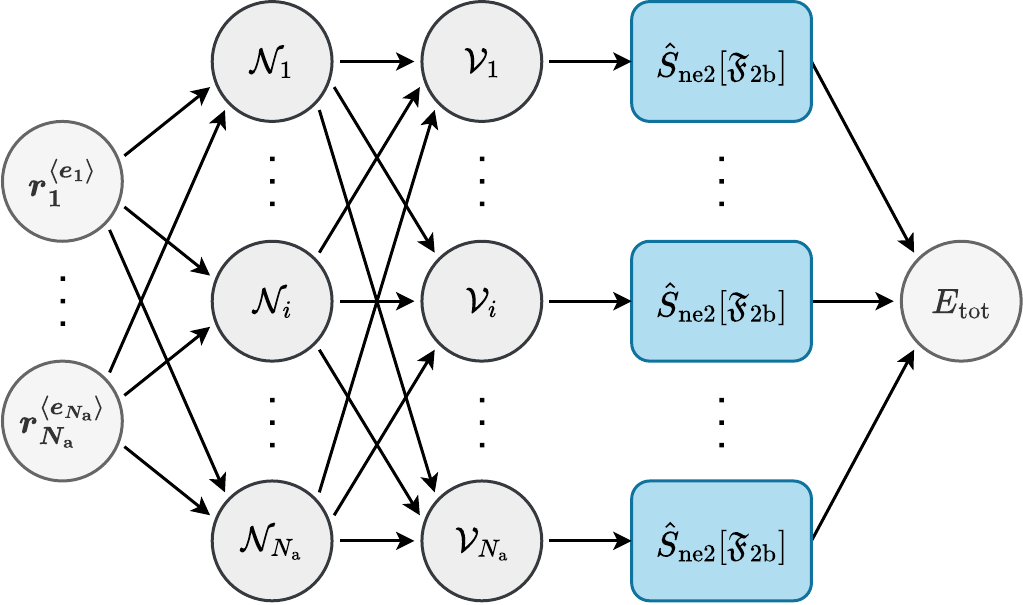}
        \caption{The computation graph of applying $\hat{S}_{{\rm ne}2}[\mathfrak{F}_{\rm 2b}]$ to compute the total energy. Compared to (a), an additional layer is inserted in the graph, which consists of the atomic-cluster neighbor lists $\mathcal{V}_i$, cf. Equation~(\ref{eq:cnl}). Each list is then passed to $\hat{S}_{{\rm ne}2}[\mathfrak{F}_{\rm 2b}]$ in the third layer, whose outputs are eventually summed up.}
        \label{fig:ne2b1}
    \end{subfigure}
    \hfill
    \begin{subfigure}[t]{0.45\textwidth}
        \centering
        \includegraphics[width=\textwidth]{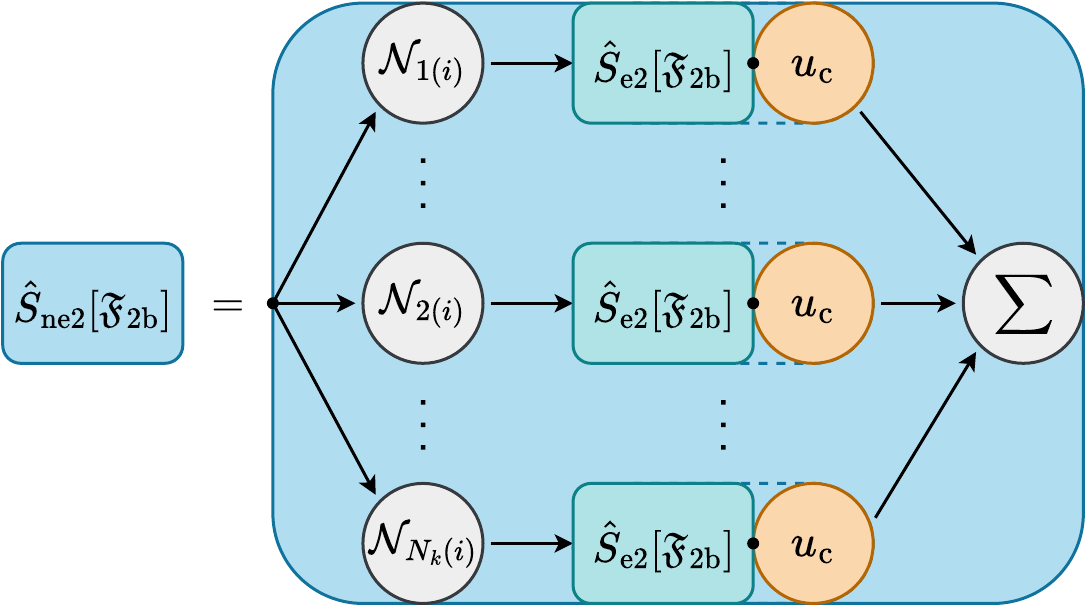}
        \caption{The computation graph of $\hat{S}_{{\rm ne}2}[\mathfrak{F}_{\rm 2b}]$. The input $\mathcal{V}_i$ is decomposed into tuple nodes $\mathcal{N}_{k(i)} \!=\!(\mathcal{N}_k,\,r_{ik})$ to form the first layer. Each tuple is then passed to a composite node in the second layer, which computes the product of $\hat{S}_{{\rm e}2}[\mathfrak{F}_{\rm 2b}](\mathcal{N}_k)$ and $u_{\rm c}(r_{ik};\,r_{\rm c})$. The outputs of the third layer are summed up as the model output.}
        \label{fig:ne2b2}
    \end{subfigure}
    \caption{The computation graphs of the two proposed nonlinear single-term models. (a) shows the overall pipeline of $\hat{S}_{{\rm e}2}[\mathfrak{F}_{{\rm 2b}}]$ and (b) shows its internal structure. The respective graphs for the neighboring-exponentiated pair-cluster interaction model $\hat{S}_{{\rm ne}2}[\mathfrak{F}_{{\rm 2b}}]$ are in (c) and (d).}\label{fig:nlst}
\end{figure*}

Like the multilayer perception (MLP) commonly used to form NNs, the exponential mapping architecture of $\hat{S}_{{\rm e}2}$ also adds nonlinear connectivity among $\psi_{2{\rm b},\nu}$ as the model descriptors. However, compared to a single-layer perception, $f(\bm{x}) \coloneqq h(\bm{c}\cdot \bm{x} + \tau)$, where $h$ is a nonlinear activation function and $\bm{c}$ are linear weights, the nonlinear mapping and inner product operation of $\hat{S}_{{\rm e}2}$ are in a reverse order, cf. Equation (\ref{eq:e2b}). In fact, the architecture of $\hat{S}_{{\rm e}2}$ can be considered a generalized application of the Kolmogorov--Arnold representation (KAR) theorem~\cite{kolmogorov1961representation, braun2009constructive}. The KAR theorem provides a formula for representing a continuous multivariate function as the composition of univariate functions. It is a theoretical backup for using the form of Equation (\ref{eq:e2b}) to approximate the high-dimensional interatomic potential with the one-dimensional two-body interaction fragments generated by the basis functions $\psi_{2{\rm b},\nu}$. The derivation of the connecting $\hat{S}_{{\rm e}2}[\mathfrak{F}_{\rm 2b}]$ back to the KAR theorem is provided in Appendix~\ref{app:kar}, and the computation graphs of this model are shown in FIG. \ref{fig:e2b1} and \ref{fig:e2b2}.

Additionally, we define an \textbf{atomic-cluster} neighbor list for the $i$th atom by
\begin{equation}\label{eq:cnl}
    \mathcal{V}_i \equiv \mathcal{N}_i  \cup \bigl\{ \mathcal{N}_k\,\big|\,k\neq i;\,\lVert\bm{r_{ik}}\rVert_2 \!\leq\! r_{\rm c}\bigr\},
\end{equation}
where the included neighbor lists $\mathcal{N}_k$ represent the atomic clusters that are in the vicinity of the $i$th atom (specified by $\mathcal{N}_i$). Replacing $\mathcal{N}_i$ with $\mathcal{V}_i$ as the input, we propose a neighboring-exponentiated pair-cluster interaction potential:
\begin{equation}\label{eq:ne2b}
    \begin{aligned}
        &\phantom{\equiv}\;E_{S_{{\rm ne}2}[\mathfrak{F}_{\rm 2b}]}\bigl(\mathcal{V}_i; \, \bm{\theta_{S_{{\rm ne}2}[\mathfrak{F}_{\rm 2b}]}}\bigr)\\
        &\!\!\equiv \sum_{k \neq i}
        u_{\rm c} \bigl( r_{ik};\,r_{\rm c} \bigr)\,
        E_{S_{{\rm e}2}[\mathfrak{F}_{\rm 2b}]} \bigl(\mathcal{N}_k;\, \bm{\theta_{S_{{\rm e}2}[\mathfrak{F}_{\rm 2b}]}}
        \bigr)\\[4pt]
        &\mathrm{with}\hspace{0.5em}\bm{\theta_{S_{{\rm ne}2}[\mathfrak{F}_{\rm 2b}]}} = \bm{\theta_{S_{{\rm e}2}[\mathfrak{F}_{\rm 2b}]}}, 
    \end{aligned}
\end{equation}
where $u_{\rm c}$ is the same cutoff function used for the many-body cluster basis functions, cf. Equation ~(\ref{eq:cutoff}) in Appendix~\ref{app:lcb}. The computation graphs of $\hat{S}_{{\rm ne}2}[\mathfrak{F}_{\rm 2b}]$ are shown in FIG.~\ref{fig:ne2b1} and \ref{fig:ne2b2}. 

Comparing FIG.~\ref{fig:e2b1} and FIG.~\ref{fig:ne2b1}, we can see that $\hat{S}_{{\rm ne}2}$ implements an additional layer before the nonlinear mappings to extract the collective effect of atomic-cluster interactions between the center atomic cluster ($\mathcal{N}_i$) and its neighbors ($\mathcal{N}_k$). This cluster-wise correlation cannot be realized solely by implementing more sophisticated MLIP models~\cite{behler2007generalized, schutt2018schnet} as each model is designed to only have access to the local information provided by one $\mathcal{N}_i$ bounded by a finite cutoff radius $r_{\rm c}$.

\section{Dual-term interatomic potentials}\label{sec:dtm}
In Sec.~\ref{sec:stm}, we have introduced three single-term MLIP model architectures ($\hat{S}_{{\rm l}2}$, $\hat{S}_{{\rm e}2}$, $\hat{S}_{{\rm ne}2}$), as well as the distinct function forms of their output interatomic potentials when acting on the basis set $\mathfrak{F}_{\rm 2b}$. In this section, we propose the simplest forms of multi-term model architectures, \textbf{dual-term} architectures. Specifically, we focus on applying summation and multiplication operations to single-term models and studying the FIM eigenspectra of the resulting composite MLIP models.

Again, we first consider the simplest case. Given two  model predictions with respect to a one-dimensional data point $x_s$: 
\begin{equation}
        q_{1,s} \coloneqq \mathcal{M}_{1}\!\left(x_{s};\,\bm{\theta_1}\right),\;
        \;
        q_{2,s} \coloneqq \mathcal{M}_{2}\!\left(x_{s};\,\bm{\theta_2}\right);
\end{equation}
based on Equation~(\ref{eq:likelihoodregression}), the NLLs for the sum and the product of them are
\begin{equation}
\begin{split}
&{\rm{NLL}}_{\mathcal{M}_1+\mathcal{M}_2}\!\left(\theta\right) = \sum_{s}w_{s}\left|p_{s} - q_{1,s} - q_{2,s}\right|^{2},\\
&{\rm{NLL}}_{\mathcal{M}_1\times \mathcal{M}_2}\!\left(\theta\right) = \sum_{s}w_{s}\left|p_{s} - q_{1,s} \times q_{2,s}\right|^{2}
 \end{split}
\end{equation}
where $\bm{\theta} = \bm{\theta_1} \cup \bm{\theta_2}$. Correspondingly, the respective FIMs for those two model compositions are
\begin{equation}\label{eq:fim_dt}
    \begin{aligned}
        \mathcal{I}^{\braket{+}}\!\!
        =\,& 2 \!\sum_{s}\!w_s\! \Bigl[\bm{V_{1,s}} \!+ \bm{V_{2,s}} \!+ \bm{U_{s}} +\\
        &\phantom{2 \!\sum_{s}\!\! w_s[}\left(q_{1,s}\!\!+\!q_{2,s} \!\!-\! p_s\right)\!\left(\bm{H_{1,s}}\!\!+\!\bm{H_{2,s}}\right)\Bigr],\\
        \mathcal{I}^{\braket{\times}}\!\!
        =\,& 2 \!\sum_{s}\!w_s\! \Bigl[q_{2,s}^2\!\bm{V_{1,s}} \!+ q_{1,s}^2\!\bm{V_{2,s}} \!+ 
        \left(q_{1,s} \!\times q_{2,s}\right)\bm{U_{s}}+\\
        &\phantom{2 \!\sum_{s}\!\! w_s [}\left(q_{1,s}\!\!\times\!q_{2,s}\!\!-\!p_s\right)\!\left(q_{2,s}\bm{H_{1,s}} \!\!+\! q_{1,s}\bm{H_{2,s}} \!\!+\!
         \bm{U_{s}}\right)\Bigr];
    \end{aligned}
\end{equation}
where
\begin{equation}\label{eq:fim_dt_sym}
    \begin{aligned}
        \bm{V_{i,s}} &= \frac{\partial q_{i,s}}{\partial \bm{\theta}}\frac{\partial q_{i,s}}{\partial \bm{\theta}^\mathsf{T}},\quad
        \bm{U_{s}} = \frac{\partial q_{1,s}}{\partial \bm{\theta}}\frac{\partial q_{2,s}}{\partial \bm{\theta}^\mathsf{T}} + \frac{\partial q_{2,s}}{\partial \bm{\theta}}\frac{\partial q_{1,s}}{\partial \bm{\theta}^\mathsf{T}},\\
        \bm{H_{i,s}} &= \frac{\partial^2q_{i,s}}{\partial\bm{\theta}\partial\bm{\theta}^\mathsf{T}}.
    \end{aligned}
\end{equation}

From Equation~(\ref{eq:fim_dt}) and (\ref{eq:fim_dt_sym}), it is clear to see that due to the presence of $\bm{V_{i,s}}$ and $\bm{U_{s}}$, the FIMs of the two composed model predictions are not simply block-diagonal matrices out of $\mathcal{I}^{\braket{M_1}}$ and $ \mathcal{I}^{\braket{M_2}}$, even if $\bm{\theta_1} \cap \bm{\theta_2} \!=\! \varnothing$. This structure suggests that the FIM of a dual-term model may have a different eigenvalue distribution than its two single-term submodels. If $M_{1}\!\left(x_{s};\,\bm{\theta_1}\right)$ has an ill-conditioned FIM, is it possible to design $M_{2}\!\left(x_{s};\,\bm{\theta_2}\right)$ such that the FIM for their dual-term composition becomes well-conditioned? To explore the possibility of using dual-term compositions to improve model accuracy while optimizing the condition number of the FIM, we present several prototypical dual-term models in the following sub-sections.

\subsection{``Term $+$ Term'' models}
First, we define a dual-model addition operator:

\begin{equation}\label{eq:t+t}
    \begin{aligned}
        &\phantom{\equiv}\;\hat{D}_{+}\!\bigl[ \hat{M}_{\rm L}\!\left[ \mathfrak{F}_{\rm L}\right],\hat{M}_{\rm R}\!\left[\mathfrak{F}_{\rm R}\right]\bigr]\\
    &\!\!\equiv\bm{x} \mapsto \hat{M}_{\rm L}\!\left[ \mathfrak{F}_{\rm L}\right]\left(\bm{x_{\rm L}}\right) + \hat{M}_{\rm R}\!\left[ \mathfrak{F}_{\rm R}\right]\left(\bm{x_{\rm R}}\right)\\
    &\mathrm{with}\hspace{0.5em} \bm{x} = \bm{x_{\rm L}} \cup \bm{x_{\rm R}}.
    \end{aligned}
\end{equation}
By consecutively using $\hat{D}_{+}$ to combine models capturing distinct physical characteristics, such as $n$-body interactions for different orders (i.e., $n$), we can approach the ground-truth interatomic potential. Herein, we introduce a dual-term architecture that combines two submodels, capturing many-body interactions of the same order but different types, as opposed to those of the same type but different orders, with respect to the center atom. Specifically, we propose an exponentiated-neighboring--exponentiated pair-cluster model:
\begin{equation}\label{eq:ene2b}
    \begin{aligned}
        &\hat{P}_{\rm ene2}[\mathfrak{F}_{\rm 2b}] \equiv \hat{D}_{+}\bigl[\hat{S}_{\, {\rm e}2}[\mathfrak{F}_{\rm 2b}], \, \hat{S}_{{\rm ne2}[+]}[\mathfrak{F}_{\rm 2b}]\bigr]\\
        & \mathrm{with}\hspace{0.5em}\bm{\theta_{P_{\rm ene2}[\mathfrak{F}_{\rm 2b}] }} =
         \bm{\theta_{S_{\rm e2}}} \,\cup\, 
         \bm{\theta_{S_{{\rm ne2}[+]}}} \,\cup\, 
         \bm{\theta_{\mathfrak{F}_{\rm 2b}}}.
    \end{aligned}
\end{equation}
Unlike the generic addition operation shown in Equation~(\ref{eq:t+t}), where a different basis set may be used for each term, the underlying dual-term architecture $\hat{P}_{\rm ene2}$ enforces the same two-body cluster basis set $\mathfrak{F}_{\rm 2b}$ for both single-term architectures, $\hat{S}_{\, {\rm e}2}$ and $\hat{S}_{\, {\rm ne2}[+]}$. This constraint of basis sets imposes a uniform resolution in capturing different types of two-body interactions, which are pair-cluster interactions in different vicinities with respect to the center atom.

\subsection{``Term $\times$ Term'' models}
Second, we define a dual-model multiplication operator:
\begin{equation}\label{eq:t*t}
    \begin{aligned}
        &\phantom{\equiv}\;\hat{D}_{\times}\!\bigl[ \hat{M}_{\rm L}\!\left[ \mathfrak{F}_{\rm L}\right],\hat{M}_{\rm R}\!\left[\mathfrak{F}_{\rm R}\right]\bigr]\\
    &\!\!\equiv\bm{x} \mapsto \hat{M}_{\rm L}\!\left[ \mathfrak{F}_{\rm L}\right]\left(\bm{x_{\rm L}}\right) \times \hat{M}_{\rm R}\!\left[ \mathfrak{F}_{\rm R}\right]\left(\bm{x_{\rm R}}\right)\\
    &\mathrm{with}\hspace{0.5em} \bm{x} = \bm{x_{\rm L}} \cup \bm{x_{\rm R}}.
    \end{aligned}
\end{equation}
Compared to $\hat{D}_{+}$, the composition of multiple $\hat{D}_{\times}$ does not directly introduce more types of interactions, but rather promotes the same type of interactions to higher orders. By utilizing this operator, we propose a bilinear pair-cluster product model:
\begin{equation}\label{eq:pp}
    \begin{aligned}
        &\hat{P}_{\, {\rm l2l2}}[\mathfrak{F}_{\rm 2b}]\equiv \hat{D}_{\times[+]}\biggl[\hat{S}_{{\rm l}2}^{\braket{\rm L}}\left[{\mathfrak{F}^{\braket{\rm L}}_{\rm 2b}}\right], \, \hat{S}_{{\rm l}2}^{\braket{\rm R}}\left[{\mathfrak{F}^{\braket{\rm R}}_{\rm 2b}}\right]\biggr]\\
        &\mathrm{with}\hspace{0.5em} \bm{\theta_{P_{\, {\rm l2l2}}[\mathfrak{F}_{\rm 2b}]}} =
        \bm{\theta_{S^{\braket{\rm L}}_{\, {\rm l2}}}} \cup 
        \bm{\theta_{S^{\braket{\rm R}}_{\, {\rm l2}}}} \cup 
        \bm{\theta_{\mathfrak{F}_{\rm 2b}}} \cup 
        \bm{\tau}^{\braket{\mathfrak{A}}}.
    \end{aligned}
\end{equation}

The direct benefit of dual-term product models, such as $\hat{P}_{\, {\rm l2l2}}[\mathfrak{F}_{\rm 2b}]$, which are composed of two single-term models, is the suppression of scaling for the parameter set size. For instance, by multiplying only two pair-interaction models, $\hat{P}_{\, {\rm l2l2}}$ gains the capability to describe three-body interactions. Although the three-body cluster basis functions $\psi_{3{\rm b},\nu}$, cf. Equation~(\ref{eq:3bcb}), can directly capture three-body interaction, a linear model composed of it requires $\mathcal{O}(M^2)$ linear coefficients given $M$ localized two-body functions $\Phi_{f_m}$. In contrast, with the same number of $\Phi_{f_m}$, $\hat{P}_{\, {\rm l2l2}}$ only requires $\mathcal{O}(M)$ model coefficients, cf. Equation~(\ref{eq:pp}). 

Finally, to introduce additional flexibility into the model while optimizing the condition number of the FIM, we relax the restriction of having the same architecture for the two submodels (terms). Specifically, we propose a linear-exponential pair-cluster product model $\hat{P}_{\, {\rm l2e2}}$:
\begin{equation}\label{eq:pe}
    \begin{aligned}
        &\hat{P}_{\, {\rm l2e2}}[\mathfrak{F}_{\rm 2b}]\equiv \hat{D}_{\times[+]}\biggl[\hat{S}_{{\rm l}2}\left[{\mathfrak{F}^{\braket{\rm L}}_{\rm 2b}}\right], \, \hat{S}_{\, {\rm e}2}\left[{\mathfrak{F}^{\braket{\rm R}}_{\rm 2b}}\right]\biggr]\\
        &\mathrm{with}\hspace{0.5em} \bm{\theta_{P_{\, {\rm l2l2}}[\mathfrak{F}_{\rm 2b}]}} =
        \bm{\theta_{S_{\, {\rm l2}}}} \cup 
        \bm{\theta_{S_{\, {\rm e2}}}} \cup 
        \bm{\theta_{\mathfrak{F}_{\rm 2b}}} \cup
        \bm{\tau}^{\braket{\mathfrak{A}}}.
    \end{aligned}
\end{equation}

For both $\hat{P}_{\, {\rm l2l2}}$ and $\hat{P}_{\, {\rm l2e2}}$, their basis set $\mathfrak{F}_{\rm 2b}(\bm{\theta_{\mathfrak{F}_{\rm 2b}}})$ is always split into two disjoint subsets ${\mathfrak{F}^{\braket{\rm L}}_{\rm 2b}}(\bm{\theta_{\rm L}})$ and ${\mathfrak{F}^{\braket{\rm R}}_{\rm 2b}}(\bm{\theta_{\rm R}})$ to be applied for the two submodel architectures, and they do not share any parameters. To summarize the architectures of the dual-term models we proposed in this section, as well as their relations to the single-term models introduced in Sec.~\ref{sec:stm}, we have compiled a summary of them in TABLE~\ref{tb:models}.
\begin{table}[htp]
    \setlength{\tabcolsep}{1.7pt}
    \renewcommand{\arraystretch}{1.6}
    \centering
    \begin{tabular}{cccrrcc}
        \toprule[1.5pt]
        \multicolumn{2}{c}{\multirow{2}{*}{Composition Type}} & \multirow{2}{*}{\shortstack{Archi-\\tecture}} & \multicolumn{3}{c}{Model Structure} & \multirow{2}{*}{Eq.} \\ \cline{4-6}
        \multicolumn{2}{c}{} &  & \multicolumn{1}{c}{\footnotesize Left-term} & \multicolumn{1}{c}{\footnotesize Right-term} & {\footnotesize Shift} &  \\ \midrule[1.25pt]
        \multirow{5}{*}{\shortstack{Single-\\term}} & Linear & $\hat{S}_{{\rm l}2}$ & \multicolumn{2}{c}{\multirow{3}{*}{N/A}} & No & (\ref{eq:l2b}) \\ \cline{2-3} \cline{6-7}
         & \multirow{4}{*}{\shortstack{Non-\\linear}} & $\hat{S}_{{\rm e}2}$ & \multicolumn{2}{c}{} & No & (\ref{eq:e2b}) \\
         &  & $\hat{S}_{{\rm ne}2}$ & \multicolumn{2}{c}{} & No & (\ref{eq:ne2b}) \\
         &  & $\hat{S}_{{\rm e2}[+]}$ & $\hat{S}_{\rm e2}[\mathfrak{F}_{\rm 2b}]$ & \multicolumn{1}{c}{-} & Yes & - \\
         &  & $\hat{S}_{{\rm ne2}[+]}$ & $\hat{S}_{\rm ne2}[\mathfrak{F}_{\rm 2b}]$ & \multicolumn{1}{c}{-} & Yes & - \\ \midrule[1.0pt]
        \multirow{3}{*}{\shortstack{Dual-\\term}} & Sum ($\hat{D}_{+}$) & $\hat{P}_{\rm ene2}$ & $\hat{S}_{{\rm e}2}[\mathfrak{F}_{\rm 2b}]$ & $\hat{S}_{{\rm ne2}[+]}[\mathfrak{F}_{\rm 2b}]$ & No & (\ref{eq:ene2b}) \\ \cline{2-7} 
         & \multirow{2}{*}{Product ($\hat{D}_{\times}$)} & $\hat{P}_{\, {\rm l2l2}}$ & $\hat{S}_{{\rm l}2}[\mathfrak{F}_{\rm L}]$ & $\hat{S}_{{\rm l}2}[\mathfrak{F}_{\rm R}]$ & Yes & (\ref{eq:pp}) \\
         &  & $\hat{P}_{\, {\rm l2e2}}$ & $\hat{S}_{{\rm l}2}[\mathfrak{F}_{\rm L}]$ & $\hat{S}_{{\rm e2}[+]}[\mathfrak{F}_{\rm R}]$ & Yes & (\ref{eq:pe}) \\
         \bottomrule[1.5pt]
    \end{tabular}
    \caption{Summary of the single-term and dual-term models proposed in Sec.\ref{sec:stm} and \ref{sec:dtm}. For each single-term model, its linearity is included; for each dual-term model, the composition of its left-term and right-term submodels is specified. ``Shift'' indicates whether an extra shifting parameter is added through the mapping defined in Equation~(\ref{eq:shift}). The last column lists the equation numbers of related model definitions.}
    \label{tb:models}
\end{table}

\section{{Fisher information matrix} guided model {development} in the case of Niobium}\label{sec:results}
To study the accuracy and stability of the single-term and dual-term models proposed in Sec.~\ref{sec:stm}--\ref{sec:dtm}, we tested their training performances against a GT dataset of niobium (Nb), which was previously also used to develop interatomic potentials based on the physics-inspired GEAM model~\cite{sun2023exploring}. As discussed in Ref.~\onlinecite{sun2023exploring}, this dataset includes bulk structures of body-centered cubic, face-centered cubic, and hexagonal close-packed phases of Nb under varying degrees of hydrostatic compression and tetragonal shear. It also includes structures with point defects, such as vacancies and different interstitial configurations, planar defects (e.g., stacking faults), symmetric-tilt grain boundaries, and liquid structures at a few temperature-pressure points. The total energies and atomic forces were obtained from DFT calculations performed by using the Vienna Ab initio Simulation Package (VASP).\cite{Kresse1993, kresse1996efficiency, kresse1996} The composition of the sample points in the dataset is summarized in TABLE~\ref{tb:dataset}.

\begin{table}[htp]
\setlength{\tabcolsep}{1.7pt}
    \renewcommand{\arraystretch}{1.2}
    \centering
    \begin{tabular}{cccc}
        \toprule[1.5pt]
        \multicolumn{4}{c}{Number of various sample points} \\
        Total  & Energy & Force & Non-zero Force  \\\midrule[1.25pt]
        112613 & 125    & 37496 ($\times$3) & 20783 ($\times$3)\\
        \bottomrule[1.5pt]
    \end{tabular}\caption{The Nb ground-truth dataset composition.}
    \label{tb:dataset}
\end{table}

The loss function $\mathcal{L}$ for training the model configurations is defined as the residual sum of squares, i.e., an equal-weight least-squares fitting based on Equation~(\ref{eq:likelihoodregression}), with respect to the GT dataset:
\begin{equation}\label{eq:loss}
    \begin{aligned}
        \mathcal{L}\left(\bm{\theta_{M[\mathfrak{F}]}}\right) \equiv \sum_s \biggl( \Big|E_{\rm tot}^{\braket{s}} -\mathcal{E}_{\rm tot}^{\braket{s}} \Big|^2 \!+\! 
        \sum_{i=1}^{N_{\rm a}^{\braket{s}}} \Big\lVert \bm{F_i}^{\braket{s}} - \bm{\mathcal{F}_i^{\braket{s}}} \Big\rVert^2_2 \biggr),
    \end{aligned}
\end{equation}
where $\mathcal{E}_{\rm tot}^{\braket{s}}$ and $\bm{\mathcal{F}_i^{\braket{s}}}$ are the GT values of the total energy and the atomic force (on the $i$th atom) in the $s$th structure, respectively. In contrast, $E_{\rm tot}^{\braket{s}}$ is the total energy of the $s$th structure predicted by the model configuration $\hat{M}[\mathfrak{F}]$ at $\bm{\theta_{M[\mathfrak{F}]}}$, cf. Equation~(\ref{eq:pes1}); $\bm{F_i}^{\braket{s}}$ is the predicted force on the $i$th atom in the same structure, obtained by evaluating its analytic form derived from the model configuration:
\begin{equation}
    \bm{F_i}^{\braket{s}} = -\frac{\partial E_{\rm tot}^{\braket{s}}}{\partial \bm{r_i^{\braket{s}}}} \approx -\frac{\partial E_{M[\mathfrak{F}],i^{\braket{s}}} \bigl(\bm{\theta_{M[\mathfrak{F}]}} \bigr)}{\partial \bm{r_i^{\braket{s}}}}.
\end{equation}

\subsection{Training procedure}\label{sec:training}
\begin{figure}[ht]
\centering
\includegraphics[width=0.45\textwidth]{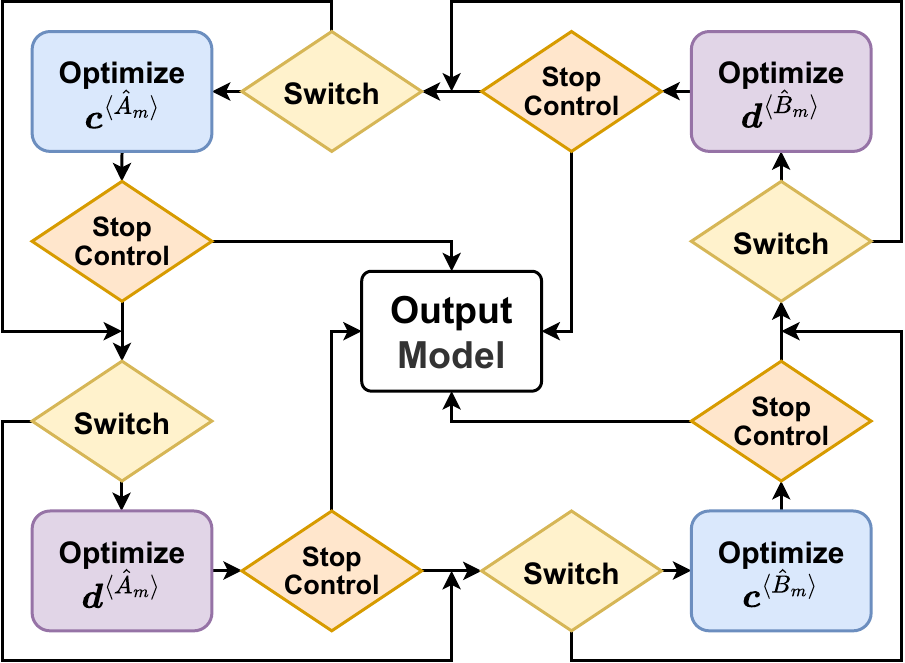}
\caption{A dual-bipartite training procedure for a decomposable model $\hat{M}[\mathfrak{F}] \coloneqq \hat{D}_m[\hat{A}_m[\mathfrak{F}_A],\, \hat{B}_m[\mathfrak{F}_B]]$. The total model parameters $\bm{\theta_{M[\mathfrak{F}]}}$ are divided into four subsets by Equation (\ref{eq:sepPars}). $\bm{c^{\braket{\hat{A}_m}}}$ ($\bm{d^{\braket{\hat{A}_m}}}$) and $\bm{c^{\braket{\hat{B}_m}}}$ ($\bm{d^{\braket{\hat{B}_m}}}$) are the linear (nonlinear) coefficients of submodels $\hat{A}_m[\mathfrak{F}_A]$ and $\hat{B}_m[\mathfrak{F}_B]$, respectively. During each training cycle, $\hat{M}[\mathfrak{F}]$ is optimized with respect to only one subset, controlled by the ``Switches'' (the yellow rhombuses). The iteration ends as soon as any ``Stop Control'' (the orange rhombuses) returns ``true,'' either because the loss function $\mathcal{L}$ has converged or the limit of training cycles has been reached.}
\label{fig:training}
\end{figure}

To train the various configurations of proposed model architectures, we apply a unified divide-and-conquer approach. First, we divide the model parameters $\bm{\theta_{M[\mathfrak{F}]}}$ into four parts by applying a bi-partition twice:
\begin{equation}\label{eq:sepPars}
    \bm{\theta_{M[\mathfrak{F}]}} \coloneqq \bigcup_{\hat{t}_m \in \{\hat{A}_m,\,\hat{B}_m\}}\!\!\!\left(\bm {c^{\braket{\hat{t}_m}}} \,\mathlarger{\cup}\, \bm{d^{\braket{\hat{t}_m}}} \right),
\end{equation}
where the first bi-partition decompose $\hat{M}[\mathfrak{F}]$ into two submodels: $\hat{M}[\mathfrak{F}] \coloneqq \hat{D}_m[\hat{A}_m[\mathfrak{F}_A],\, \hat{B}_m[\mathfrak{F}_B]]$; the second bi-partition divides the parameters of each submodel into its linear coefficients $\bm{c^{\braket{\hat{t}_m}}}$ (including the shifting parameters) and nonlinear coefficients $\bm{d^{\braket{\hat{t}_m}}}$. Since the construction of multi-term models is iteratively done through dual-model operators, the decomposition procedure is equally straightforward (see TABLE~\ref{tb:models}). To optimize the parameters of each submodel, we used linear regression to fit the linear coefficients and Bayesian optimization with sequential domain reduction techniques~\cite{BOdomainReduction2002, bayeOptLib} to optimize the nonlinear coefficients~\cite{gardner2014bayesian}. The flowchart of this \textbf{dual-bipartite} training procedure is shown in FIG.~\ref{fig:training}. 

Due to the modular separations of the parameter optimization, this training procedure allows on-the-fly adjustments, such as filtering tunable parameters or switching optimizers for more efficient traversal on the high-dimensional landscape of the loss function. In the case of training single-term models, the models are treated as the first part (the submodel $\hat{A}_m$) in the training pipeline, with the second part being skipped by the corresponding ``Switches'' (the yellow rhombuses in FIG.~\ref{fig:training}).

\subsection{Evaluation method}\label{sec:eval}
To evaluate the performance of the proposed model architectures in describing various physical characteristics of the Nb system, we employed a set of training error metrics based on four types of root mean square errors (RMSE). They represent four specific properties, respectively: energy, force, force amplitude, and force angle. The corresponding formulae are
\begingroup
\allowdisplaybreaks
\begin{align}
    {\rm RMSE}_{\bar{E}} &\!=\!\! \left(\frac{\sum_s^{N_{\!s}} \bigl(E_{\rm tot}^{\braket{s}}-\mathcal{E}_{\rm tot}^{\braket{s}}\bigr)^2/\bigl({N_{\rm a}^{\braket{s}}} \bigr)^2}{N_s}\right)^{\!\!1/2};\\
    {\rm RMSE}_{\bm{F}} &\!=\!\! \left(\frac{\sum_{s,\,i}^{N_{\!s},N_{\!\rm a}^{\braket{s}}}\!\bigl\lVert\bm{F_i}^{\braket{s}}-\bm{\mathcal{F}_i}^{\braket{s}}\bigr\rVert_2^2}{\sum_s^{N_s}3N_{\rm a}^{\braket{s}}}\right)^{\!\!1/2};\\
    {\rm RMSE}_{\lVert\bm{F}\rVert} &\!=\!\! \left(\frac{\sum_{s,\,i}^{N_{\!s},N_{\!\rm a}^{\braket{s}}}
    \!\!\bigl(
    \bigl\lVert \bm{F_i}^{\braket{s}} \bigr\rVert_2 - \bigl\lVert \bm{\mathcal{F}_i}^{\braket{s}} \bigr\rVert_2
    \bigr)^2}{\sum_s^{N_s}N_{\rm a}^{\braket{s}}}\right)^{\!\!1/2};\\
    {\rm RMSE}_{\measuredangle \bm{F}} &\!=\!\! \left(\frac{\sum_{s,\,i}^{N_{\!s},N_{\!\rm a}^{\braket{s}}}\hspace{-0.6em}
    f_{\lVert\bm{0}\rVert}\!\bigl(\bm{\mathcal{F}_i^{\braket{s}}}\bigr)
    f_{\measuredangle}\!
    \bigl(\bm{F_i}^{\braket{s}}\!\!\!,\,\bm{\mathcal{F}_i}^{\braket{s}}\bigr)
    }{\sum_{s,\,i}^{N_{\!s},N_{\!\rm a}^{\braket{s}}} 
    f_{\lVert\bm{0}\rVert}\!\bigl(\bm{\mathcal{F}_j^{\braket{s}}}\bigr)
    }\right)^{\!\!1/2}
\end{align}
\endgroup
where
\begin{equation}\label{eq:delta}
    \begin{aligned}
        f_{\lVert\bm{0}\rVert}(\bm{x}) &= \begin{cases}
        1 & \quad \text{for}\;\; \lVert\bm{x}\rVert_2 \neq 0,\\
        0 & \quad \text{for}\;\; \lVert\bm{x}\rVert_2 = 0;
    \end{cases}\\[5pt]
    f_{\measuredangle}(\bm{x},\,\bm{y}) &= \arccos^2
    \left(\!
    \frac{\bm{x}\cdot\bm{y}}
    {\rVert\bm{x}\rVert_2\lVert\bm{y}\rVert_2}
    \!\right).
    \end{aligned}
\end{equation}

In addition to the model training accuracy, we are interested in the numerical stability of different model configurations. According to the definition of the Fisher information in Equation~(\ref{eq:fim}), we also constructed the FIM with respect to the converged linear coefficients of each model $\hat{M}[\mathfrak{F}]$~\cite{machta2013parameter}:
\begin{equation}
    \mathcal{I}^{\braket{\hat{M}[\mathfrak{F}]}}_{ij} \coloneqq \frac{1}{N_s}\frac{\partial^2\mathcal{L}\bigl(\bm{\theta_{M[\mathfrak{F}]}}\bigr)}{
    \partial {c_i}^{\braket{\hat{M}[\mathfrak{F}]}}\,
    \partial {c_j}^{\braket{\hat{M}[\mathfrak{F}]}}}.
\end{equation}
 We only focus on the linear coefficients $\bm{c}$, as the optimization of the nonlinear coefficients in composite models is turned off during the later stages of model training.

As explained in Sec.~\ref{sec:lgp}, an MLIP model's FIM eigenspectrum behaves as an indicator of its numerical stability (sloppiness) with respect to a least-squares loss function. By combining the analysis of the FIM with the benchmarking of the four error metrics, we correlate the error distributions of various model configurations to the sloppiness of the MLIP architectures proposed in Sec. \ref{sec:stm} and \ref{sec:dtm}. In the following two subsections, we shall demonstrate the effectiveness of such an FIM-guided model evaluation method for finding the optimal architecture that balances model bias and numerical stability. To ensure an accurate estimation of the FIM eigenspectra and condition numbers, all FIM eigenvalues were computed using a 256-bit floating-point number system based on the GNU MPFR Library~\cite{fousse2007mpfr}.

\subsection{Results: Single-term models}\label{sec:singleTerm}
We proposed both linear and nonlinear single-term models in Sec.~\ref{sec:stm} as the components for composing models of more complex architectures. Thus, we first estimate their individual performance and stability to establish a benchmark baseline for dual-term models.

\subsubsection{$\hat{S}_{\,\rm l2}$ based models}
The first type of single-term models we examined is the pair-interaction models based on $\hat{S}_{\,\rm l2}$, which forms the linear combination of the two-body cluster basis functions $\psi_{2{\rm b},\nu}$ (plus a shifting parameter). We compared the effect of different implementations of the two-body cluster basis sets $\mathfrak{F}_{\rm 2b}$ by the respective training performance of $\hat{S}_{\,\rm l2}[\mathfrak{F}_{\rm 2b}]$. Specifically, we tested the version based on (even-tempered) Gaussian functions ($\mathfrak{F}_{{\rm 2b},\mathcal{G}}$) against the one based on the Chebyshev polynomials ($\mathfrak{F}_{{\rm 2b},\mathcal{C}}$). The construction of the $\mathcal{G}$-based (i.e., Gaussian-type) two-body cluster basis functions is presented by Equation~(\ref{eq:et}--\ref{eq:gtmbc}). To construct the corresponding $\mathcal{C}$-based version, we first define the corresponding primitive function as
\begin{equation}
\mathcal{C}_m(r_{ij};\,\{r_{\rm c}\}) \equiv T_m\bigl( g_{\rm norm}(r_{ij}, r_{\rm c}) \bigr),
\end{equation}
where $T_m$ is the $m$th-order Chebyshev polynomial of the first kind, and $g_{\rm norm}$ is a normalization function:
\begin{equation}
    g_{\rm norm}(r_{ij};  r_{\rm c}) = \begin{cases}
        (1\!-\!2r_{ij}/r_{\rm c})^3 & \quad \text{for}\;\; r_{ij}\!<\!r_{\rm c},\\
        -1                           & \quad \text{for}\;\; r_{ij}\!\geq\!r_{\rm c}.
    \end{cases}
\end{equation}
Similar to the form of Equation~(\ref{eq:bf1}), we then construct the localized two-body functions based on Chebyshev polynomials:
\begin{equation}\label{eq:bf2}
    \Phi_{{\mathcal{C}_m}} \bigl( \bm{r_{ij}};\, \varnothing\bigr) \equiv \mathcal{C}_m \bigl( r_{ij};\,\{r_{\rm c}\} \bigr)\,u_{\rm c} \bigl( r_{ij};\,r_{\rm c} \bigr).
\end{equation}
$\Phi_{{\mathcal{C}_m}}$ has no trainable parameters (indicated by $\varnothing$). As a result, the $\mathcal{C}$-based pair-interaction model $\hat{S}_{\,\rm l2}[\mathfrak{F}_{{\rm 2b},\mathcal{C}}]$ only contains linear coefficients. The performance differences between these two types of basis sets for $\hat{S}_{\,\rm l2}$ (G$\nu$ versus C$\nu$) are shown in the first 6 rows of TABLE~\ref{table:singleTermPartial}. 

\begin{table}[ht]
    \setlength{\tabcolsep}{1.8pt}
    \renewcommand{\arraystretch}{1.25}
    \centering
    \begin{tabular}{lrrrrrrr}
        \toprule[1.5pt]
        \multicolumn{1}{l}{\multirow{2}{*}{Configuration}} & \multicolumn{2}{c}{Size} & \multicolumn{1}{c}{\multirow{2}{*}{$\lg\kappa$}} & \multicolumn{4}{c}{RMSE} \\ \cline{2-3}\cline{5-8}
        \multicolumn{1}{c}{} & \multicolumn{1}{c}{$\bm{c}$} & \multicolumn{1}{c}{$\bm{d}$} & \multicolumn{1}{c}{} & \multicolumn{1}{c}{$\measuredangle \bm{F}$} & \multicolumn{1}{c}{$\lVert\bm{F}\rVert$} & \multicolumn{1}{c}{$\bm{F}$} & \multicolumn{1}{c}{$\bar{E}$} \\ \midrule[1.25pt]
        G8               & 9  & 2  & 28.38 & 1.472 & 0.674 & 0.988 & 0.108  \\
        G16              & 17 & 2  & 24.80 & 1.636 & 0.374 & 0.487 & 0.060  \\
        G18              & 19 & 2  & 29.03 & 1.664 & 0.367 & 0.479 & 0.060  \\\midrule[1.0pt]
        C8               & 9  & 0  & 12.31 & 1.652 & 0.387 & 2.140 & 0.302  \\
        C32              & 33 & 0  & 31.12 & 1.889 & 0.460 & 0.667 & 0.078  \\
        C40              & 41 & 0  & 31.66 & 1.965 & 0.479 & 0.652 & 0.072  \\\midrule[1.0pt]
        E[{\small G4L4}] & 17 & 6  & 27.77 & 1.564 & 0.415 & 0.444 & 0.035  \\
        E[{\small G4L8}] & 33 & 10 & 31.64 & 1.563 & 0.364 & 0.366 & 0.030  \\
        E[{\small G8L4}] & 33 & 6  & 25.78 & 1.484 & 0.304 & 0.262 & 0.017  \\
        \midrule[1.0pt]
        N[{\small E[{\fs G8L4}]}] & 33 & 6  & 24.82 & 1.594 & 0.466 & 0.315 & 0.018 \\
        \bottomrule[1.5pt]
    \end{tabular}
    \caption{The training results of single-term models based on different basis sets and architectures. Specifically, $\hat{S}_{\,\rm l2}$ with Gaussian-type two-body cluster basis sets $\mathfrak{F}_{{\rm 2b},\mathcal{G}}$ (G8, G16, G18) versus with Chebyshev-type two-body cluster basis sets $\mathfrak{F}_{{\rm 2b},\mathcal{C}}$ (C8, C32, C40), $\hat{S}_{{\rm e2}[+]}[\mathfrak{F}_{{\rm 2b},\mathcal{G}}]$ (in the following three rows), and $\hat{S}_{{\rm ne2}[+]}[\mathfrak{F}_{{\rm 2b},\mathcal{G}}]$ (N[{\small E[{\fs G8L4}]}]). The error metrics of each model configuration are RMSEs of four aspects: the energy $\bar{E}$ ({\evpatm}), the force $\bm{F}$ ({\evpang}), the force amplitude $\lVert\bm{F}\rVert$ ({\evpang}), and the force angle $\measuredangle \bm{F}$ (unitless), respectively. $\bm{c}$ represents the number of linear coefficients, and $\bm{d}$ represents the number of nonlinear coefficients. $\lg \kappa$ represents the logarithmic condition number of the linear-coefficient Fisher information matrix ($\mathcal{I}$).}
    \label{table:singleTermPartial}
\end{table}

Compared to the $\hat{S}_{\,\rm l2}[\mathfrak{F}_{{\rm 2b},\mathcal{C}}]$, the accuracy of $\hat{S}_{\,\rm l2}[\mathfrak{F}_{{\rm 2b},\mathcal{G}}]$ converged faster with respect to the number of basis functions used in the model. More significantly, $\hat{S}_{\,\rm l2}[\mathfrak{F}_{{\rm 2b},\mathcal{G}}]$ achieved better RMSEs for both force and energy per atom. This suggests that the even-tempered Gaussian functions may be better suited for spanning the local interatomic potential surface than the Chebyshev polynomials. This difference in the performances of the two basis sets can also be attributed to the cutoff function $u_{\rm c}$ used in this work, cf. Equation~(\ref{eq:cutoff}). $u_{\rm c}$ is equal to one when the bond length is zero and smoothly decays to zero when the bond length is equal to the cutoff radius of the model. The second and third derivatives of the cutoff function also converge to zero at the cutoff distance. Since the product of the cutoff function with Chebyshev polynomials is not an orthogonal basis set, many basis functions are needed to model the two-body interactions. We note that by using a step-function-type cutoff function, the two-body interactions can be modeled by using fewer Chebyshev polynomials~\cite{lindsey2017chimes}. However, for such a cutoff function, the higher-order derivatives at the cutoff distance are not zero.

\begin{figure}[ht]
    \centering
    \begin{subfigure}[b]{0.49\textwidth}
        \centering
        \includegraphics[width=\textwidth]{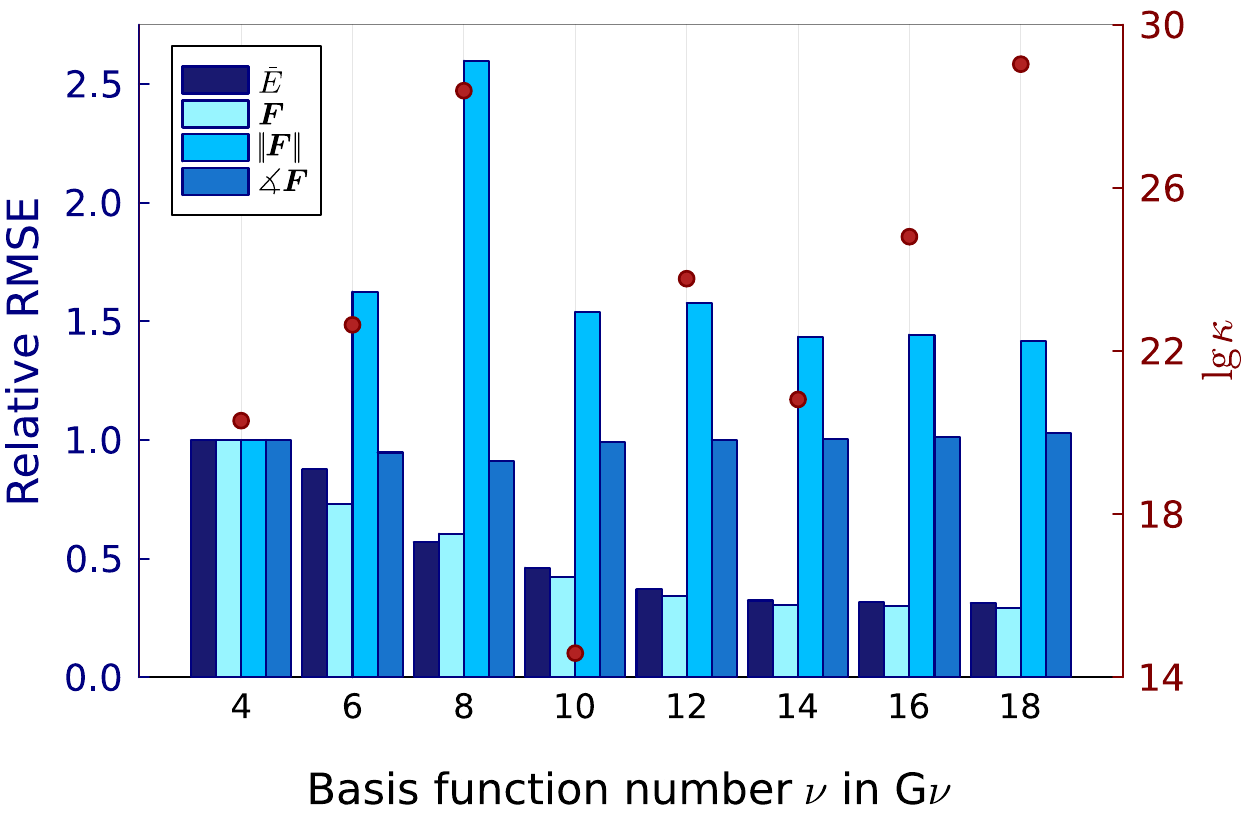}
        \caption{The four error metrics and the logarithmic condition number ($\lg \kappa$) of the linear-coefficient Fisher information matrix ($\mathcal{I}$) for each $\hat{S}_{\,\rm l2}$ configuration.}
        \label{fig:LGsP1}
    \end{subfigure}
    \begin{subfigure}[b]{0.49\textwidth}
        \centering
        \includegraphics[width=\textwidth]{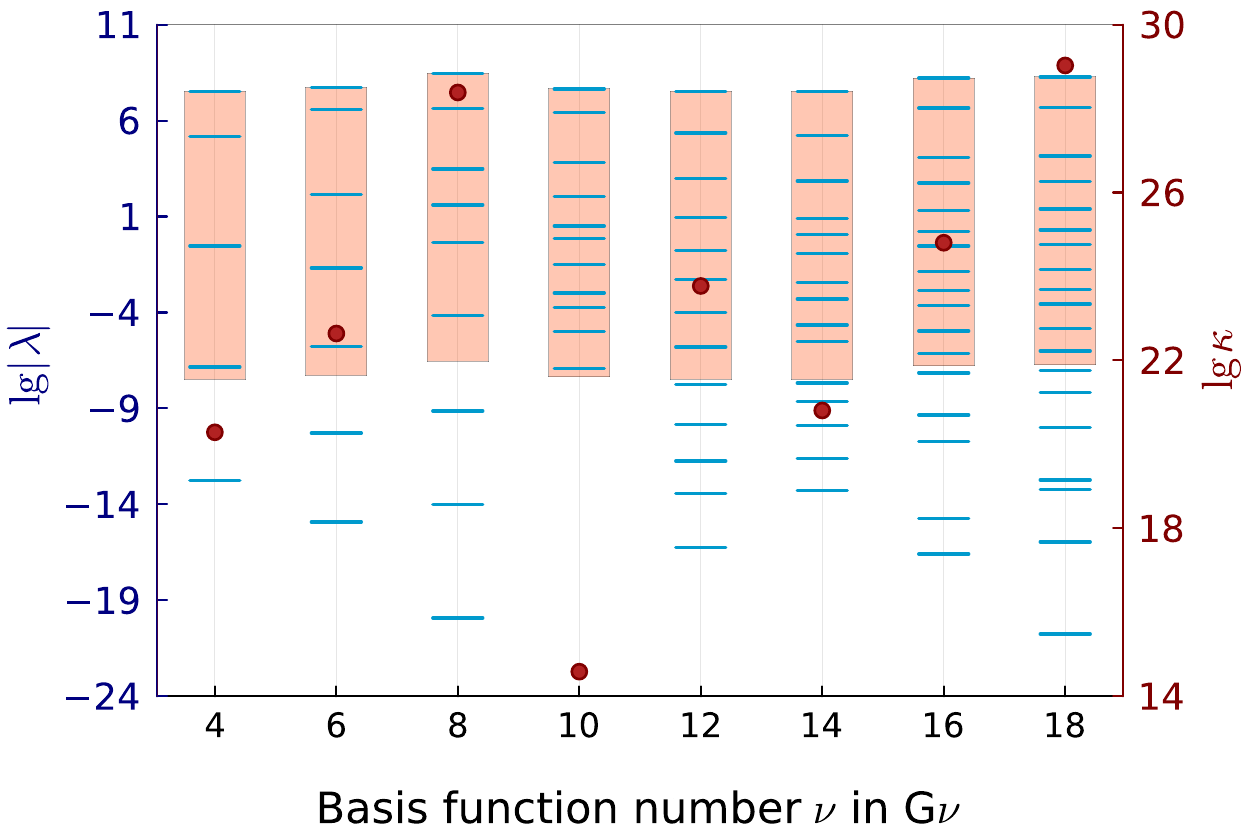}
        \caption{The logarithmic eigenspectra ($\lg |\lambda|$) of $\mathcal{I}$ corresponding to the models shown
        in FIG.~\ref{fig:LGsP1}. The orange boxes enclose the eigenvalues representing ``significant correlations (SC)''. Each of them encloses a spectrum spanning 15 orders of magnitude from the largest eigenvalue.}
        \label{fig:LGsP2}
    \end{subfigure}
    \caption{Accuracy and stability ($\lg |\lambda|$) of $\hat{S}_{\,\rm l2}$ with Gaussian-type two-body cluster basis set $\mathfrak{F}_{{\rm 2b},\mathcal{G}}$ of different sizes.}
    \label{fig:LGs}
\end{figure}

Focusing on using the more performant $\mathfrak{F}_{{\rm 2b},\mathcal{G}}$ as the default basis set version, we further tested more configurations of $\hat{S}_{\,\rm l2}$. FIG.~\ref{fig:LGsP1} summarizes the root mean squared model errors and condition numbers of the Fisher information matrices, $\kappa$, for eight different model configurations with increasing basis set size. Overall, $\kappa$ increased with respect to the basis set size, except when the optimal basis set size was achieved by using 10 even-tempered Gaussian basis functions (G10). In that case, $\kappa$ dropped to the lowest value.  Nevertheless, the force-angle ($\measuredangle \bm{F}$) and force-amplitude ($\lVert\bm{F}\rVert$) RMSEs of G10 were still worse than those of the initial model configuration, G4. This indicates the innate incapability of $\hat{S}_{\,\rm l2}$ to capture the angular (orientational) information of the interatomic interactions, as the basis functions $\psi_{2{\rm b},\nu}$ only encode the scalar interatomic distances. As we further extended $\hat{S}_{\,\rm l2}$  from G10 by adding more basis functions, the improvement of the error metrics diminished while $\kappa$ increased, reflecting deterioration of the stability of $\hat{S}_{\,\rm l2}$. This further suggests that $\kappa$ can reflect the saturation of basis functions applied to a composable model. Therefore, one should choose the model configuration with the minimal basis set as it provides better stability with similar errors compared to other basis set configurations.

An especially noteworthy case is $\hat{S}_{\,\rm l2}$ with 12 Gaussian-type cluster basis functions: the value of $\kappa$ for G12 was especially high. The speculated reason behind this observation is that the training loss function $\mathcal{L}$ is not a properly regularized measure of the model's distance to its converged state. As $\hat{S}_{\,\rm l2}[\mathfrak{F}_{{\rm 2b},\mathcal{G}}]$ is being trained (i.e., $\mathcal{L}$ is still decreasing), it reaches a sub-manifold in the functional space where only the partial errors regarding specific properties of the ground truth (e.g., total energy) have converged to a minimum. Further training does not reduce the overall errors of the model. Instead, it redistributes them such that the RMSEs for certain physical properties decrease at the cost of the RMSEs for other properties increasing. For instance, in FIG.~\ref{fig:LGsP1}, as the number of basis functions increased from four to eight, ${\rm RMSE}_{\lVert\bm{F}\rVert}$ increased whilst other RMSEs decreased. Similar error redistribution happened again when the configuration of $\hat{S}_{\,\rm l2}$ changed from G10 to G12. To further support this conjecture, we compared the eigenspectrum of the linear FIM $\mathcal{I}$ for different basis set sizes, as shown in FIG.~\ref{fig:LGsP2}. The eigenspectrum of $\mathcal{I}$ ranks the significance of different combinations among the model's linear parameters after training. Thus, we can define a region of ``significant correlations (SC)'' where the eigenvalues are at most 15 orders of magnitude smaller than the largest one. This lower-bound ratio, $10^{-15}$, is chosen to correspond approximately to the machine precision of a double-precision (64-bit) floating point number. 

As the basis set extends, on the one hand, the number of $\mathcal{I}$ eigenvalues increases. On the other hand, the eigenspectrum tends to widen (resulting in an increase of $\kappa$) unless the basis set size achieves an optimal value. These two effects compete to affect the number of eigenvalues inside the SC region. The computed FIMs of converged models met our expectations: G4, G6, G8, and G12 all have fewer eigenvalues inside the SC region than G10. Particularly, the eigenspectra of G8 and G12 form a roughly uniform distribution in the logarithmic scale, a typical characteristic of numerical models with high sloppiness~\cite{machta2013parameter, transtrum2015perspective}.

By evaluating the training performance of $\hat{S}_{\,\rm l2}$ with different basis set sizes, we show a systematic correlation between its accuracy and stability. Noticeably, such a numerical analysis guided by property-oriented error metrics and $\mathcal{I}$ eigenspectrum is not exclusive to $\hat{S}_{\,\rm l2}$, It can also be used for systematic investigation of the relation between performance and uncertainty of other MLIP architectures proposed in this paper.

\subsubsection{$\hat{S}_{\rm e2}$ based models}
To overcome the limited expressibility of $\hat{S}_{\,\rm l2}$ and optimize the systemic bias of single-term models, we proposed the exponentiated pair-cluster interaction architecture $\hat{S}_{\rm e2}$ in Sec.~\ref{sec:nlp}. This nonlinear single-term architecture encodes the collective interatomic correlations using the linear combination of outputs from a latent space characterized by the two-body interactions. To demonstrate the effectiveness of this extra parameter space, we trained several model configurations based on $\hat{S}_{\rm e2}$, and their training accuracies are also included in TABLE~\ref{table:singleTermPartial}. Specifically, each E[G$\nu$L$\mu$] represents a $\hat{S}_{{\rm e2}[+]}$ with $\nu$ Gaussian-type two-body cluster basis functions and $\mu$ latent space layers. Compared to the linear combinations of the basis functions in $\hat{S}_{\,\rm l2}$, the latent space of $\hat{S}_{{\rm e2}[+]}$ significantly improves the utilization of the basis sets, achieving a higher expressivity. For instance, using the same basis-set type and size as G8, but with the extra four latent layers, E[G8L4] obtained lower ${\rm RMSE}_{\bar{E}}$, ${\rm RMSE}_{\bm{F}}$, and $\kappa$. It even outperformed G18 in all four error metrics.

\begin{figure}[htp]
    \centering
    \begin{subfigure}[b]{0.49\textwidth}
        \centering
        \includegraphics[width=\textwidth]{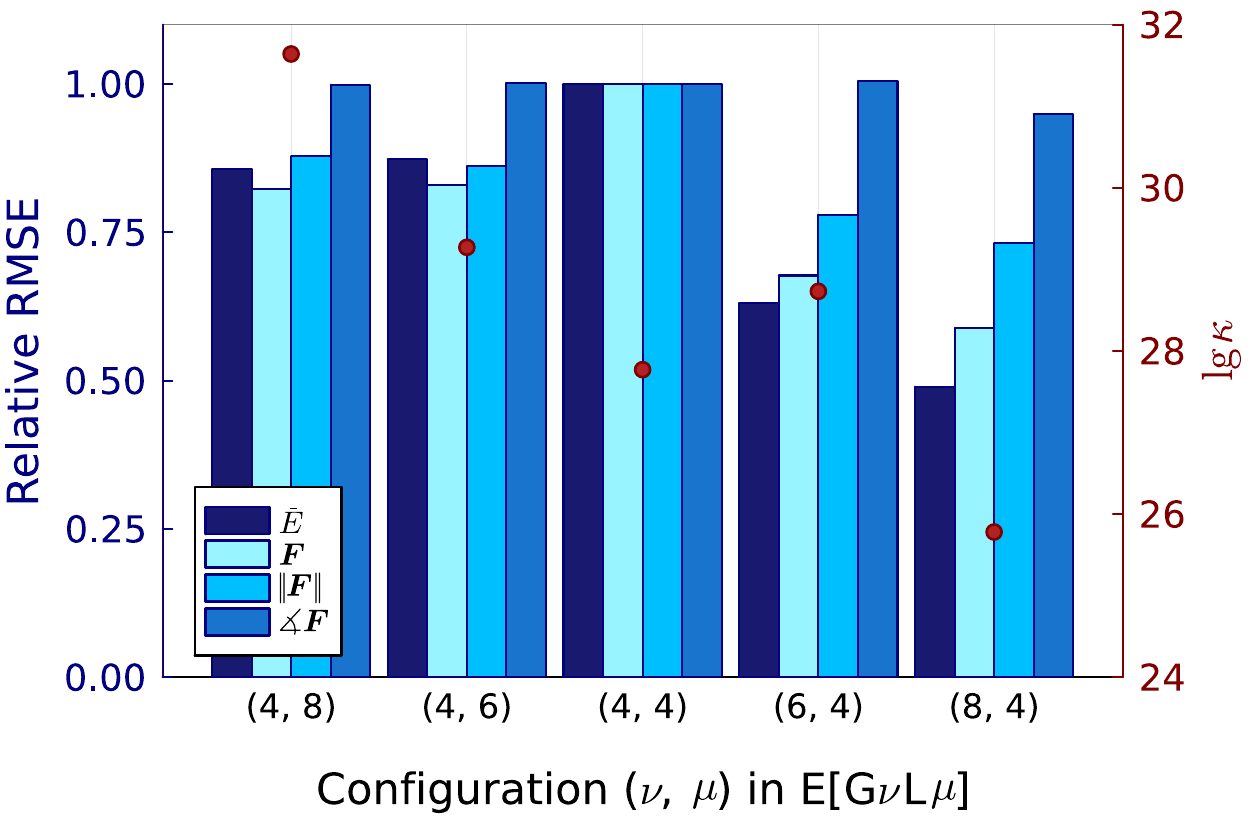}
        \caption{The four error metrics and the logarithmic condition number ($\lg \kappa$) of the linear-coefficient Fisher information matrix ($\mathcal{I}$) for each $\hat{S}_{\rm e2}$ based model configuration.}
        \label{fig:LEGsP1}
    \end{subfigure}
    \begin{subfigure}[b]{0.49\textwidth}
        \centering
        \includegraphics[width=\textwidth]{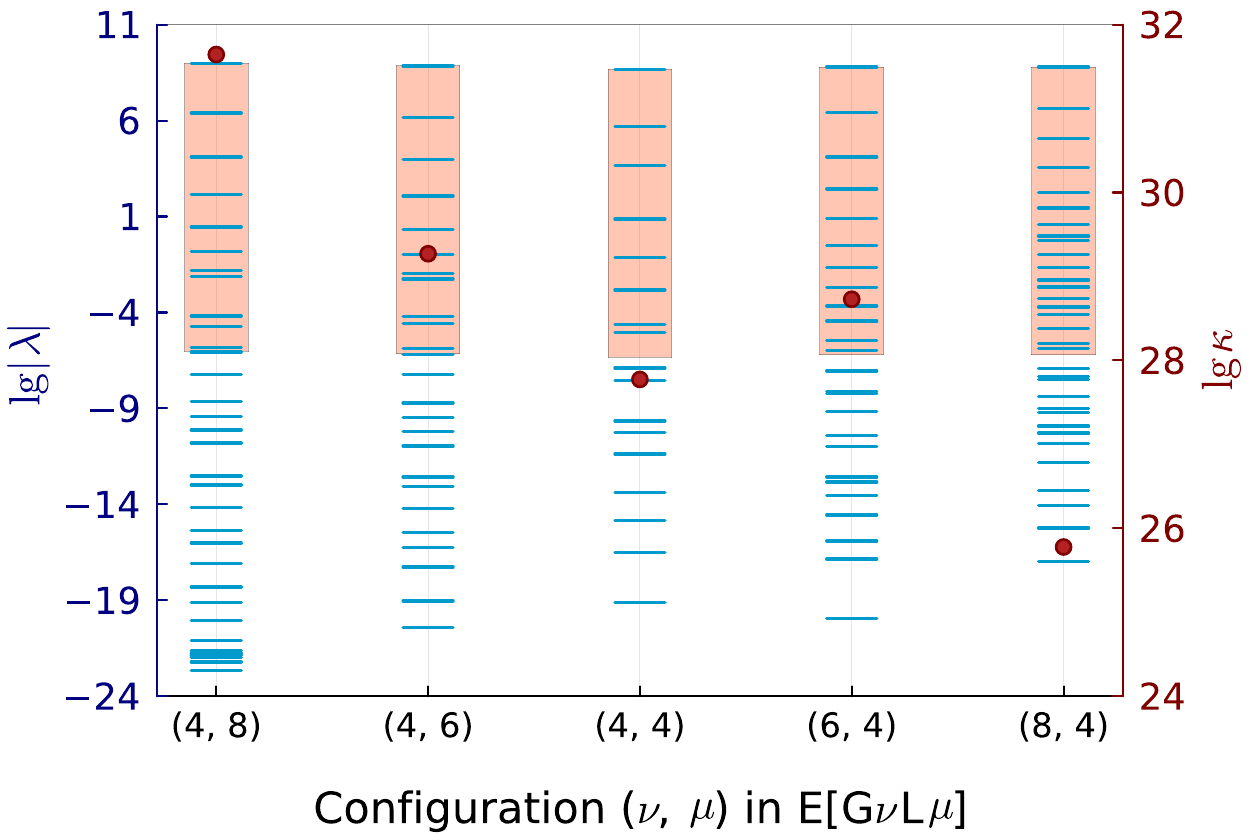}
        \caption{The logarithmic eigenspectra ($\lg |\lambda|$) of $\mathcal{I}$ corresponding to the models shown
        in FIG.~\ref{fig:LEGsP1}. The orange boxes enclose the eigenvalues representing ``significant correlations (SC)''. Each of them encloses a spectrum spanning 15 orders of magnitude from the largest eigenvalue.}
        \label{fig:LEGsP2}
    \end{subfigure}
    \caption{Accuracy and stability ($\lg |\lambda|$) of $\hat{S}_{\rm e2}$ with Gaussian-type two-body cluster basis set $\mathfrak{F}_{{\rm 2b},\mathcal{G}}$ of different sizes.}
    \label{fig:LEGs}
\end{figure}

To investigate the optimal model configuration for the basis-set size ($\nu$) and the latent-space layer number ($\mu$), we plotted the four error metrics and the $\mathcal{I}$ eigenspectra for $\hat{S}_{\rm e2}$ in FIG.~\ref{fig:LEGs}. Compared to the baseline configuration E[G4L4], for which four basis functions and four latent space layers are used, both adding basis functions and layers improved $\hat{S}_{\rm e2}$'s performance, with the contribution from the increasing basis-set size being more significant. The eigenspectra for the different configurations shown in FIG.~\ref{fig:LEGsP2} further suggest that adding latent-space layers can include more eigenvalues into the SC region.

\subsubsection{$\hat{S}_{{\rm ne2}}$ based models}
For the neighboring-exponentiated pair-cluster interaction model, which shares the same kernel structure as $\hat{S}_{{\rm e2}}$ while having a different input encoding scheme (see FIG.~\ref{fig:nlst}), we did not perform an exhaustive configuration-based training. Instead, we compared the performance of the best model configuration of $\hat{S}_{{\rm e2}[+]}$ (i.e., E[{\small G8L4}]) against a comparable configuration of $\hat{S}_{{\rm ne2}[+]}$ (i.e., 
N[{\small E[{\fs G8L4}]}]). In particular, the two model configurations have the same number of linear and nonlinear coefficients, as well as the same type of basis sets. The training result of N[{\small E[{\fs G8L4}]}] is shown in the last row of TABLE~\ref{table:singleTermPartial}. As expected, the errors of N[{\small E[{\fs G8L4}]}] were not as low as E[{\small G8L4}], since the interatomic interactions within a cluster are typically stronger than those among the neighboring clusters. However, due to the higher expressivity of the nonlinear parameterization, N[{\small E[{\fs G8L4}]}] still outperformed the linear single-term models.

\subsection{Results: Dual-term models}\label{sec:dualTerm}
In Sec.~\ref{sec:dtm}, we introduced the dual-term interatomic potential architectures that combine two single-term models with a dual-model operator. In this subsection, we show the training results of these simplest composite model architectures using only $\mathfrak{F}_{{\rm 2b},\mathcal{G}}$. TABLE~\ref{table:dualTermPartial} shows the four error metrics for various model configurations based on $\hat{P}_{\, {\rm l2l2}}$ (G4$\times$G4 and G8$\times$G8), $\hat{P}_{\, {\rm l2e2}}$ (G8$\times$E[{\small G4L4}], G8$\times$E[{\small G4L8}], G8$\times$E[{\small G8L4}], and G10$\times$E[{\small G8L4}]), and $\hat{P}_{\rm ene2}$ (E[{\small G6L5}]$+$N[{\small E[{\fs G6L5}]}] and E[{\small G8L4}]$+$N[{\small E[{\fs G8L4}]}]).

\begin{table}[htp]
    \setlength{\tabcolsep}{1.8pt}
    \renewcommand{\arraystretch}{1.25}
    \centering
    \begin{tabular}{lrrrrrrr}
        \toprule[1.5pt]
        \multicolumn{1}{l}{\multirow{2}{*}{Configuration}} & \multicolumn{2}{c}{Size} & \multicolumn{1}{c}{\multirow{2}{*}{$\lg\kappa$}} & \multicolumn{4}{c}{RMSE} \\ \cline{2-3}\cline{5-8}
        \multicolumn{1}{c}{} & \multicolumn{1}{c}{$\bm{c}$} & \multicolumn{1}{c}{$\bm{d}$} & \multicolumn{1}{c}{} & \multicolumn{1}{c}{$\measuredangle \bm{F}$} & \multicolumn{1}{c}{$\lVert\bm{F}\rVert$} & \multicolumn{1}{c}{$\bm{F}$} & \multicolumn{1}{c}{$\bar{E}$} \\ \midrule[1.25pt]
        G4  $\times$ G4                 & 11 & 4  & 29.64 & 1.573 & 0.308 & 1.325 & 0.169 \\
        G8  $\times$ G8                 & 19 & 4  & 31.34 & 1.568 & 0.329 & 0.316 & 0.028 \\\midrule[1.0pt]
        G8  $\times$ E[{\small G4L4}]   & 27 & 8  & 29.82 & 1.490 & 0.316 & 0.258 & 0.017 \\
        G8  $\times$ E[{\small G4L8}]   & 43 & 12 & 32.89 & 1.480 & 0.307 & 0.239 & 0.015 \\
        G8  $\times$ E[{\small G8L4}]   & 43 & 8  & 31.15 & 1.517 & 0.242 & 0.203 & 0.015 \\
        G10 $\times$ E[{\small G8L4}]   & 45 & 8  & 28.30 & 1.504 & 0.245 & 0.187 & 0.013 \\\midrule[1.0pt]
        E[{\small G6L5}] $\!\!+\!\!$ N[{\small E[{\fs G6L5}]}] & 61 & 12 & 28.16 & 1.473 & 0.207 & 0.181 & 0.013 \\
        E[{\small G8L4}] $\!\!+\!\!$ N[{\small E[{\fs G8L4}]}] & 65 & 10 & 23.97 & 1.460 & 0.198 & 0.172 & 0.013 \\\bottomrule[1.5pt]
    \end{tabular}
    \caption{The training results of composable model configurations based on dual-term architectures. The symbol ``$\times$'' represents two (left and right) single-term submodels combined by a dual-model multiplication $\hat{D}_{\times}$ to form the resulting dual-term model. Similarly, ``$+$'' represents to a dual-model addition operation using $\hat{D}_{+}$. The error metrics of each model configuration are RMSEs of four aspects: the energy $\bar{E}$ ({\evpatm}), the force $\bm{F}$ ({\evpang}), the force amplitude $\lVert\bm{F}\rVert$ ({\evpang}), and the force angle $\measuredangle \bm{F}$ (unitless), respectively. $\bm{c}$ represents the number of linear coefficients, and $\bm{d}$ represents the number of nonlinear coefficients. $\lg \kappa$ represents the logarithmic condition number of the linear-coefficient Fisher information matrix ($\mathcal{I}$).}
    \label{table:dualTermPartial}
\end{table}

\subsubsection{$\hat{P}_{\, {\rm l2l2}}$ based product models}
Compared to the results of $\hat{S}_{\,\rm l2}[\mathfrak{F}_{{\rm 2b},\mathcal{G}}]$ in TABLE~\ref{table:singleTermPartial}, multiplying two $\hat{S}_{\,\rm l2}[\mathfrak{F}_{{\rm 2b},\mathcal{G}}]$ together to form a $\hat{P}_{\, {\rm l2l2}}[\mathfrak{F}_{{\rm 2b},\mathcal{G}}]$ achieved better training accuracy than the naive sum of them (which still forms a $\hat{S}_{\,\rm l2}[\mathfrak{F}_{{\rm 2b},\mathcal{G}}]$). Particularly, all four error metrics of G8$\times$G8 (the second row of TABLE~\ref{table:dualTermPartial}) are lower than their respective parts of G16 despite having a similar number of parameters. This contrast suggests that by multiplying two $\hat{S}_{\,\rm l2}[\mathfrak{F}_{{\rm 2b},\mathcal{G}}]$ submodels, the basis functions within each submodel are utilized more sufficiently.
\begin{figure}[htp]
    \centering
    \begin{subfigure}[b]{0.49\textwidth}
        \centering
        \includegraphics[width=\textwidth]{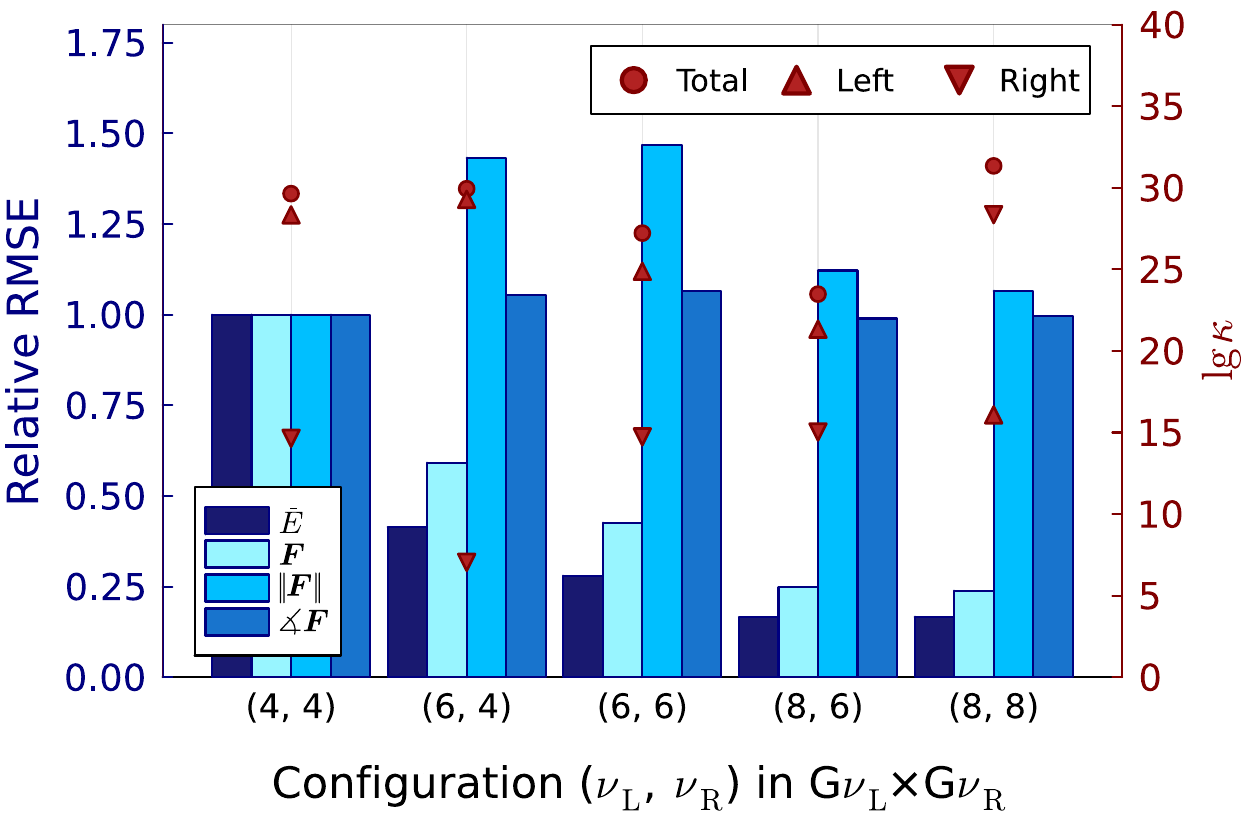}
        \caption{$\hat{P}_{\, {\rm l2l2}}$}
        \label{fig:GGs}
    \end{subfigure}
    \begin{subfigure}[b]{0.49\textwidth}
        \centering
        \includegraphics[width=\textwidth]{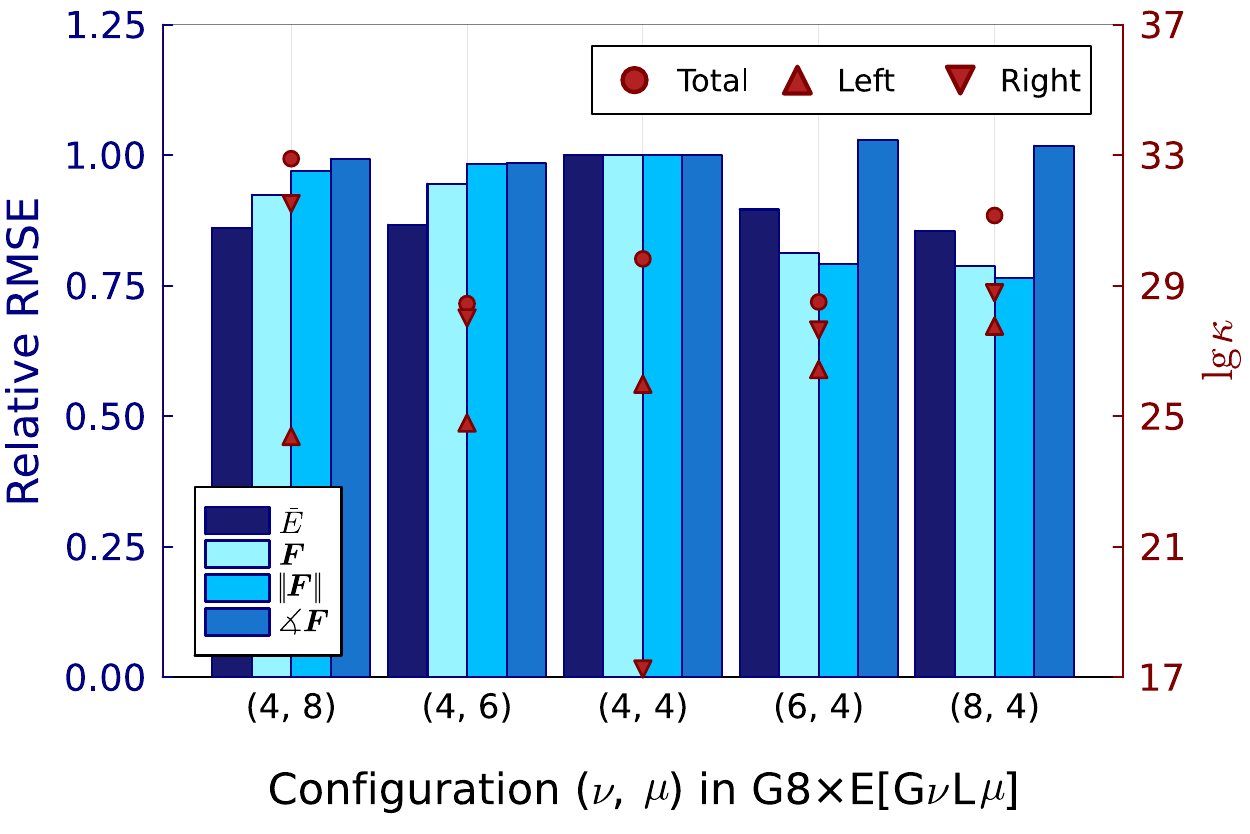}
        \caption{$\hat{P}_{\, {\rm l2e2}}$}
        \label{fig:GEGs}
    \end{subfigure}
    \caption{The four error metrics and the logarithmic condition number ($\lg \kappa$) of the linear-coefficient Fisher information matrix ($\mathcal{I}$) for each configuration of two dual-term product models, $\hat{P}_{\, {\rm l2l2}}$ and $\hat{P}_{\, {\rm l2e2}}$. $\lg\kappa$ of the product model configuration (Total), its left-term submodel (Left), and its right-term submodel (Right) are included as red markers with different shapes and guided by the right axis.}
    \label{fig:dualTermProduct}
\end{figure}
To further understand the reason behind this benefit, we plotted the error metrics for the different models in FIG.~\ref{fig:GGs}. We set the baseline configuration of $\hat{P}_{\, {\rm l2l2}}$ to be G4$\times$G4, where each submodel term is constructed with four basis functions. As the size of the basis size grew, $\kappa$ of the left-term submodel decreased, and $\kappa$ of the right-term submodel increased. Such a divergence between the stabilities of the two terms can be explained by the training procedure introduced in Sec.~\ref{sec:training}. The training of a composable potential model always starts with training one part of it. For instance, in the case of $\hat{P}_{\, {\rm l2l2}}$, the coefficients in the left-term $\hat{S}_{\, {\rm l2}}$ with $\nu_{\rm L}$ basis functions (specified by G$\nu_{\rm L}$) was first obtained by ignoring the right-term submodel. As the training of the right-term $\hat{S}_{\, {\rm l2}}$ started, it complemented the left term by learning the finer details of the ground-truth interatomic potential embedded in the training set, which gets refined as both terms are optimized iteratively. The correlation between $\kappa$ and the error metrics for single-term models presented in Sec.~\ref{sec:singleTerm} indicates that the term (submodel) in $\hat{P}_{\, {\rm l2l2}}[\mathfrak{F}_{{\rm 2b},\mathcal{G}}]$ with higher $\kappa$ learned more information from the training set at the cost of higher uncertainties along sloppy directions. Noticeably, the $\kappa$ of the dual-term product models are lower bounded by the higher $\kappa$ of their two submodels. This conclusion need not be limited to $\hat{P}_{\, {\rm l2l2}}$ since the $\kappa$--error correlations also appeared for $\hat{P}_{\, {\rm l2e2}}$ shown in FIG.~\ref{fig:GEGs}. In practice, one can check the stability of a dual-term product model during the training simply by tracking the local $\kappa$ of its submodels. This approach is less costly than constantly tracking $\kappa$ of the whole model.

\subsubsection{$\hat{P}_{\, {\rm l2e2}}$ based product models}
Although both $\hat{P}_{\, {\rm l2l2}}$ and $\hat{P}_{\, {\rm l2e2}}$ are constructed by applying the multiplication operator $\hat{D}_{\times}$, $\hat{P}_{\, {\rm l2e2}}$ uses two different sub-models and hence has a structural asymmetry across its sub-models as opposed to $\hat{P}_{\, {\rm l2l2}}$. FIG.~\ref{fig:GEGs} shows that by changing the hyperparameters $m$ and $n$, G8$\times$E[G$n$L$m$] can transit into two distinct regimes: $n>m$ and $n<m$, where the four error metrics have different distributions. However, $\hat{P}_{\, {\rm l2l2}}$ does not hold such property, cf. FIG.~\ref{fig:GGs}.

\begin{figure}[htp]
    \centering
    \begin{subfigure}[b]{0.49\textwidth}
        \centering
        \includegraphics[width=\textwidth]{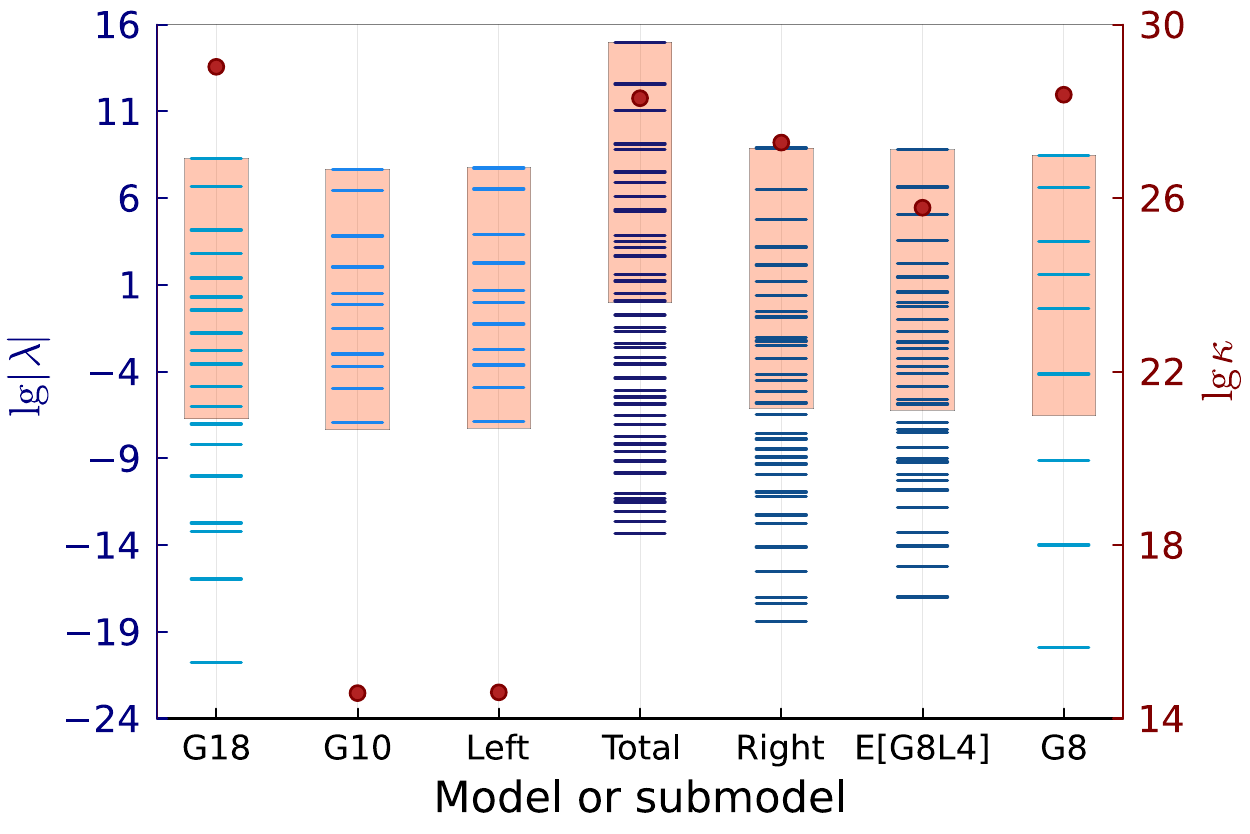}
        \caption{$\hat{P}_{\, {\rm l2e2}}$ configuration: G10$\times$E[G8L4]}
        \label{fig:G10E8L4EigVal}
    \end{subfigure}
    \begin{subfigure}[b]{0.49\textwidth}
        \centering
        \includegraphics[width=\textwidth]{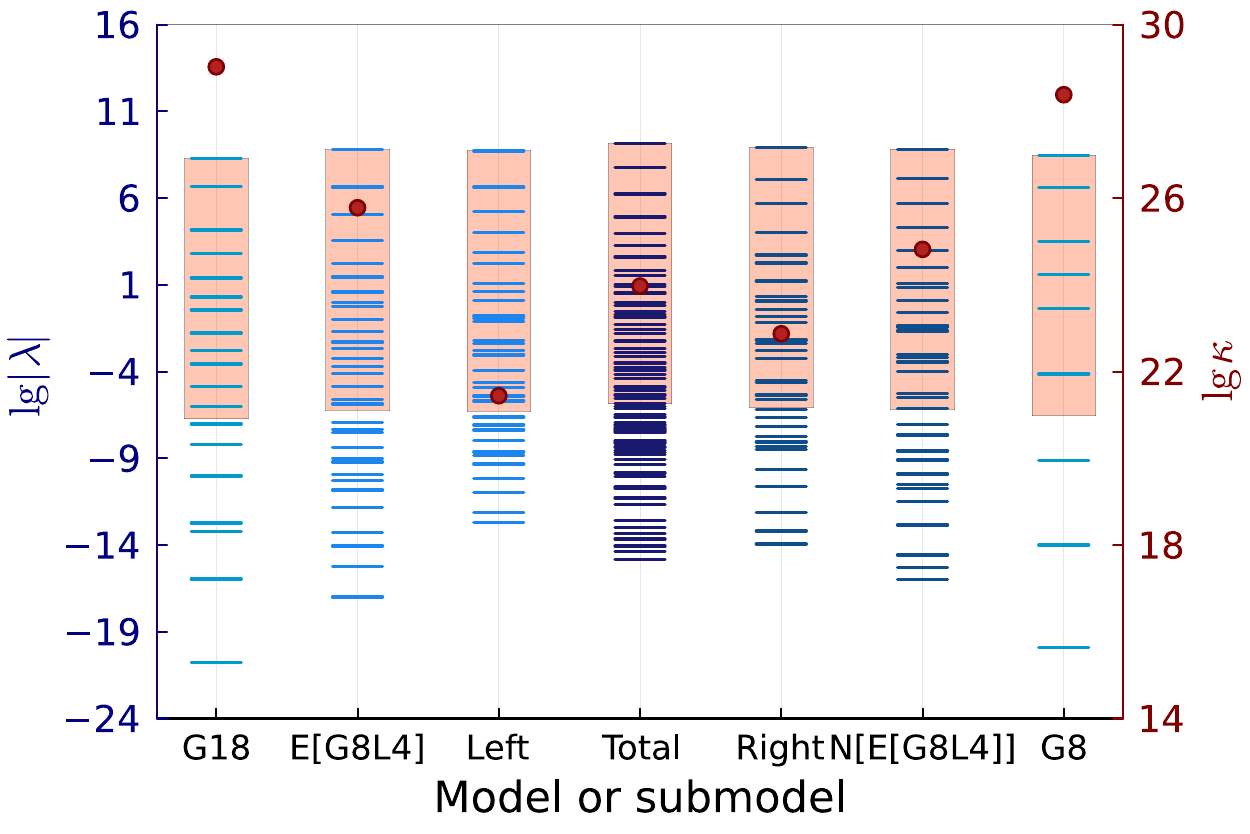}
        \caption{$\hat{P}_{\rm ene2}$ configuration: E[{\small G8L4}]$+$N[{\small E[{\fs G8L4}]}]}
        \label{fig:LENE8L4EigVal}
    \end{subfigure}
    \caption{
    The logarithmic eigenspectra ($\lg |\lambda|$) of the linear-coefficient Fisher information matrix ($\mathcal{I}$) for two composable and optimally configured dual-term potential models, compared with the relevant model (and its left-term and right-term submodel) configurations. The orange boxes enclose the eigenvalues representing ``significant correlations (SC)''. Each of them encloses a spectrum spanning 15 orders of magnitude from the largest eigenvalue. The respective $\lg\kappa$ of different models (submodels) are included as red markers and guided by the right axis.}
    \label{fig:dualTermEigVals}
\end{figure}

To further examine how optimally configured single-term submodels affect the performance of the composite potential model they become part of, we trained the model configuration, G10$\times$E[G8L4]. G10 and E[G8L4] are the configurations with the lowest $\kappa$ for $\hat{S}_{\,\rm l2}$ and $\hat{S}_{{\rm e2}[+]}$, respectively.
As a result,  G10$\times$E[G8L4] achieved the lowest $\kappa$ and one of the lowest errors across all the dual-term product models we have tested. Please refer to TABLE~\ref{table:dualTerm} in Appendix~\ref{app:data} for the complete results of all the $\hat{P}_{\, {\rm l2e2}}$ configurations that have been tested. Moreover, FIG.~\ref{fig:G10E8L4EigVal} shows the comparison between the $\mathcal{I}$ eigenspectrum of G10$\times$E[G8L4] and several relevant model (submodel) configurations. First, the product of the two single-term submodels has a noticeable shifting effect that moves the $\mathcal{I}$ eigenspectrum upwards. Second, the SC regions of the two submodels went through different degrees of transition as they were combined to form the resulting configuration G10$\times$E[G8L4]. The $\mathcal{I}$ eigenspectrum of G10 remained almost identical before and after the composition, whereas E[G8L4] changed drastically. Nonetheless, the $\mathcal{I}$ eigenspectrum inside the SC region of G10$\times$E[G8L4] is not a simple superposition of the $\mathcal{I}$ eigenspectrum for each term. This discrepancy indicates that new correlations between the two submodels were formed after they were multiplied together. This observation matches our expectation that a dual-term product model can capture higher-order many-body interactions beyond its submodel terms.

\subsubsection{$\hat{P}_{\rm ene2}$ based sum models}
Last but not least, FIG.~\ref{fig:LENE8L4EigVal} shows the $\mathcal{I}$ eigenspectrum of E[G8L4]$+$N[E[G8L4]], which has the best performance out of all the dual-term potential model configurations presented in TABLE~\ref{table:dualTermPartial}. Unlike the product architectures, $\hat{P}_{\, {\rm l2l2}}$ and $\hat{P}_{\, {\rm l2e2}}$, we observed significant compression of the $\mathcal{I}$ eigenspectrum for $\hat{P}_{\rm ene2}$. Even though the $\kappa$ of E[G8L4]$+$N[E[G8L4]] is still larger than those of its left and right terms, it is smaller than those of both E[G8L4] and N[E[G8L4]] as standalone models. By adding $\hat{S}_{{\rm ne2}[+]}$ (N[E[G8L4]]) to $\hat{S}_{\rm e2}$ (E[G8L4]) with the same two-body cluster basis sets, cf. Equation (\ref{eq:ene2b}), we managed to improve both the accuracy and the stability of $\hat{S}_{\rm e2}$. It should be stressed that the performance boost did not come from the mere inclusion of N[E[G8L4]] as one term inside the dual-model summation. This is because N[E[G8L4]] had worse performance than E[G8L4] (see TABLE~\ref{table:singleTermPartial}). The complementary contributions from both $\hat{S}_{\rm e2}$ and $\hat{S}_{{\rm ne2}[+]}$ together provides the boost of $\hat{P}_{\rm ene2}$. This conclusion is also 
 supported by the change of the $\mathcal{I}$ eigenspectrum's distribution (in addition to the compression) before and after these two single-term models were combined as the left and right terms of E[G8L4]$+$N[E[G8L4]].

\section{Conclusions and outlook}\label{sec:co}
In this paper, we proposed an adaptive strategy for physics-inspired machine-learning interatomic potential (MLIP) design based on Fisher information matrix (FIM) guided analysis and composable model architectures (FIG.~\ref{fig:adaptivedesign}). We formulated a model composition framework where the basic components for constructing the potential models are ``single-term'' models based on many-body cluster basis functions. By iteratively adjusting the model configuration based on the property-oriented error metrics and the linear-coefficient FIM eigenspectrum of the training results, we systematically improve the model design in terms of both numerical accuracy and stability.

In Sec.~\ref{sec:stm}, we introduced the single-term models and how they can be used as the building blocks for composable models with more complex structures through dual-model operations. We adapted the conventional model architectures based on $n$-body atomistic interactions to define linear single-term models $\hat{S}_{\, {\rm l}n}$. Additionally, we proposed two nonlinear single-term models based on the exponentiated pair-cluster interaction architecture $\hat{S}_{{\rm e}2}$, cf. Equation~(\ref{eq:e2b}), and the neighboring-exponentiated pair-cluster interaction architecture $\hat{S}_{{\rm ne}2}$, cf. Equation~(\ref{eq:ne2b}). These nonlinear architectures both introduce a latent parameter space spanned by an additional set of nonlinear parameters, the exponent coefficients $\bm{\zeta}$. The linear parameters encode the interatomic interactions within an atomic cluster, while the nonlinear parameters encode the interatomic interactions among the multiple clusters within the vicinity of a cluster center. Aside from proposing the single-term models and their underlying architectures, we also explained their connections to the FIM, which can be used to quantify the numerical stability of MLIP models.

By defining the dual-model addition  $\hat{D}_{+}$ and multiplication  $\hat{D}_{\times}$ operators acting on the introduced single-term models, we then proposed three dual-term composable potential architectures in Sec~\ref{sec:dtm}, $\hat{P}_{\rm ene2}$ that forms sum models, as well as $ \hat{P}_{\, {\rm l2l2}}$ and $\hat{P}_{\, {\rm l2e2}}$ that form product models. The compositions of these architectures are either based on physical inspirations for atomistic many-body interactions or numerical strategies for improving the model's FIM condition number. The connections between the proposed single-term and dual-term architectures are summarized in TABLE~\ref{tb:models}.

Finally, we showed the performance of these model architectures in relation to their single-term model components and underlying basis functions in Sec.~\ref{sec:results}. To evaluate both the accuracy and the stability of a certain model, we used an evaluation method combining the error metrics of four property-oriented RMSEs (energy, force, force amplitude, and force angle), along with the eigenspectrum of the FIM. We applied this method to analyze the training results of various model configurations against a data set of 125 niobium (Nb) structures. The results demonstrated that we are able to systematically adapt and improve the fitting performance of multiple model architectures. Ultimately, we obtained an optimal dual-term sum model configuration (E[G8L4] $\!+\!$ N[E[G8L4]] with 75 parameters) that has a force RMSE of 0.172~{\evpang} and an energy RMSE of 0.013~{\evpatm}. 

Furthermore, using the same evaluation method, we characterized the correlations between model components (single-term submodels), basis sets, and the resulting composable potential model configurations. Particularly, both the dual-term product models and the sum models presented nontrivial improvement over their submodels. However, they have different advantages. The product architecture is more extensible and compatible with a wider range of single-term models. Even in the case of combining two models of the same type, it can still outperform the naive model additions (e.g., G8$\times$G8 outperformed G16). On the contrary, for the sum architecture, a complementary submodel design is required to offer a significant performance boost. One straightforward way we have shown is deploying distinctive feature embedding. For instance, $\hat{S}_{\rm ne2}[\mathfrak{F}_{{\rm 2b},\mathcal{G}}]$ has the same internal physics-inspired structure as $\hat{S}_{\rm e2}[\mathfrak{F}_{{\rm 2b},\mathcal{G}}]$ but focuses on a different range of collective two-body interactions. When combining them together to form the sum model $\hat{P}_{\rm ene2}[\mathfrak{F}_{{\rm 2b},\mathcal{G}}]$, we managed to achieve improvement in both performance and stability, cf. TABLE~\ref{table:dualTermPartial} and FIG.~\ref{fig:dualTermEigVals}.

In the future, our composable interatomic potential design framework can be extended and improved further. Due to the clear separation between different modules within the framework (see FIG.~\ref{fig:adaptivedesign}), such extensions can be readily incorporated. First, we can explore additional modifications of the model architecture, such as acting on new many-body basis variants constructed from different types of localized two-body functions. We already showed in Sec.\ref{sec:singleTerm} that different types of localized two-body functions, e.g., Gaussian-based versus Chebyshev-polynomial-based, can sensitively impact both the accuracy and the stability of the resulting models. Several prior studies on the performance of different types of two-body functions for MLIP design provide paths for inspiration~\cite{seko2015first, takahashi2017conceptual}. We can adopt those functions to form new kinds of many-body cluster basis sets and test them against the current Gaussian-based ones. Secondly, we can investigate the limit of composable potential models beyond dual-term architectures. Since a composable potential model can always be decomposed into its two (immediate) submodels, we can recursively apply the same evaluation method proposed in this paper to compose and optimize such ``higher-order'' models. We leave this to future work. Additionally, we can combine the FIM eigenspectrum analysis with other uncertainty quantification methods, such as Cram\'er--Rao bound~\cite{kurniawan2022bayesian}, to implement a more comprehensive uncertainty quantification of different submodel composition formalisms. Last but not least, we can incorporate the model architectures proposed in this paper into other MLIP models and test the combined hybrid models on a broader range of atomistic systems.

We hope that our work in this paper will invite a new perspective on general MLIP design. Instead of starting with a complex and difficult-to-train model and working to simplify its structure, we propose beginning with simple models with clear physical motivations and building complexity iteratively. Then, by monitoring both the accuracy (error metrics) and model stability (FIM eigenspectrum), optimal model configurations with necessary expressivity are adaptively constructed and developed in an efficient manner, while maintaining a balance between model stability and extensibility.

\begin{acknowledgments}
    W.W. is grateful to Haoyuan Shi for the valuable discussions about general modeling and optimization techniques for interatomic potentials during the early stage of this research. W.W. also would like to thank Liming Zhao for helping perform numerical validations of several model configurations proposed in the paper using an independent code base.
    This work was performed under the auspices of the U.S. Department of Energy by Lawrence Livermore National Laboratory under Contract DE-AC52-07NA27344, funded by the Laboratory Directed Research and Development Program at LLNL under project tracking code 23-SI-006. LLNL review and release number is LLNL-JRNL-2005209.
\end{acknowledgments}

\bibliography{main}

\appendix

\section{Many-body cluster basis functions and sets}\label{app:lcb}
Consider a smooth scalar function parameterized by $\bm{\theta_m}$
\begin{equation}
    f_m \bigl(r_{ij};\, \bm{\theta_m} \bigr):\,\mathbb{R}\to\mathbb{R}   
\end{equation}
as one of the primitive functions to approximate any arbitrary two-body interaction. For each $f_m$, the input variable 
\begin{equation}
    r_{ij} \equiv \lVert\bm{r_{ij}}\rVert_2 = \left\lVert\bm{r_i^{\braket{e_i}}} - \bm{r_j^{\braket{e_j}}}\right\rVert_2
\end{equation}
is the distance between (the nuclei of) the $i$th atom of element $e_i$ at $\bm{r_i}$ and the $j$th atom of element $e_j$ at $\bm{r_j}$. The output value of $f_m$ represents a fragment of the two-body potential energy between the $i$th atom and $j$th atom.

To impose a local approximation of the two-body interaction, we multiply $f_m$ with a cutoff function $u_{\rm c}$ to form a localized two-body function
\begin{equation}\label{eq:l2bf}
    \Phi_{f_m} \left( \bm{r_{ij}}; \, \bm{\theta_m}\right) \equiv 
    f_m \bigl( r_{ij}; \, \bm{\theta_m} \bigr) \, u_{\rm c}\bigl( r_{ij};\,r_{\rm c} \bigr),
\end{equation}
such that $\Phi_{f_m}$ has a smooth convergence to zero at $r_{\rm c}$ which is the cutoff radius. Specifically, we choose $u_{\rm c}$ to have the following expression:
\begin{equation}\label{eq:cutoff}
    u_{\rm c} ( r;\,r_{\rm c} ) = \begin{cases}
        (1-r/r_{\rm c})^4 & \quad \text{for}\;\; r\!<\!r_{\rm c},\\
        0                 & \quad \text{for}\;\; r\!\geq\!r_{\rm c}.
    \end{cases}
\end{equation}

Letting $f_m$ be a Gaussian function 
\begin{equation}
    f_m \bigl( r_{ij};\,\bm{\theta_m}\bigr) \coloneqq \mathcal{G}_m \bigl( r_{ij};\,\{\beta_m\} \bigr) \equiv e^{-\beta_m r_{ij}^2}
\end{equation}
parameterized by an exponent coefficient $\beta_m \!>\! 0$, a common option for descriptors in interatomic potential model design~\cite{bartok2010gaussian, bartok2015g}. Subsequently, we define a $\mathcal{G}$-based (``Gaussian-type'') localized two-body function as
\begin{equation}\label{eq:bf1}
    \Phi_{{\mathcal{G}_m}} \bigl( \bm{r_{ij}};\,\{\beta_m\} \bigr) \equiv \mathcal{G}_m \bigl( r_{ij};\,\{\beta_m\} \bigr)\,u_{\rm c} \bigl( r_{ij};\,r_{\rm c} \bigr).
\end{equation}

To further encode the chemical information of the local atomistic interaction, we define a sequence of unique $N$-body chemical composition sets $\mathfrak{C}_N$, whose elements represent all possible elemental compositions specified by the target atomic symbols. For instance, considering a system composed of two elements X and Y, i.e., $\mathfrak{A} \coloneqq \{{\rm X}, {\rm Y}\}$. The generated $\mathfrak{C}_N$ are 
\begin{equation}
    \begin{aligned}
        \mathfrak{C}_2(\{\rm X, Y\}) &= \{\rm \braket{X, X}, \braket{X, Y}, \braket{Y, Y}\},\\
        \mathfrak{C}_3(\{\rm X, Y\}) &= \{\rm \braket{X,\!X,\!X}\!, \braket{X,\!X,\!Y}\!, \braket{X,\!Y,\!Y}\!, \braket{Y,\!Y,\!Y}\},\\
                      &\cdots
    \end{aligned}
\end{equation}
where $\braket{\cdot, \dots, \cdot}$ are permutation-invariant lists with the identities: 
\begin{equation}
    \begin{aligned}
        \braket{a,\!b} & \equiv \braket{b,\!a}, \\
        \braket{a,\!b,\!c} \equiv \braket{a,\!c,\!b} \equiv \braket{b,\!a,\!c} &\equiv \braket{b,\!c,\!a} \equiv \braket{c,\!a,\!b} \equiv \braket{c,\!b,\!a},\\ 
        &\dots
    \end{aligned}
\end{equation}

After generating all the atomic lists from a given $\mathfrak{C}_N$, we use these lists to label fitting coefficients associated with different types of $N$-body atomic interactions. Specifically, we define a chemical-composition-aware parameter set
\begin{equation}\label{eq:labeledEq}
    {\bm{\theta}}^{\braket{\mathfrak{C}_N(\mathfrak{A})}} \equiv 
    \bigcup_{\braket{\cdot} \in \mathfrak{C}_N(\mathfrak{A})} \bm{\theta}^{\braket{\cdot}}
\end{equation}
which is the union of all the parameter sets that are of the same type and labeled by the atomic lists generated by $\mathfrak{C}_N$. Note that $\bm{\theta}$ can be replaced by any other set-like symbols (in bold font). With ${\bm{\theta}}^{\braket{\mathfrak{C}_N(\mathfrak{A})}}$, we can form a chemistry-aware basis for a cluster of two-body atomistic interactions. Specifically, we define the general expression of the two-body cluster basis function as:
\begin{equation}\label{eq:2bcb}
        \hat{V}_{\rm 2b}[f]\left(\mathcal{N}_i;{\bm{\omega}}^{\braket{\mathfrak{C}_2(\mathfrak{A})}}\!,\bm{\theta_m} \right)
        \equiv \sum_{j\neq i} \omega^{\braket{e_i, e_j}}\Phi_{\!f_m} \!\left(\bm{r_{ij}};\bm{\theta_m}\right), 
\end{equation}
where $\mathcal{N}_i$ is the atomic neighbor list defined in Equation~(\ref{eq:neighbor_list}), and ${\bm{\omega}}^{\braket{\mathfrak{C}_2(\mathfrak{A})}}$ is a set of linear chemical-composition-aware parameters. Each parameter $\omega^{\braket{e_i, e_j}} \!\in\! {\bm{\omega}}^{\braket{\mathfrak{C}_2(\mathfrak{A})}}$ is characterized by $e_i$ and $e_j$, the atomic symbols of $i$th and $j$th atoms, respectively. Hence, $\hat{V}_{\rm 2b}[f]$, as a basis function, outputs a weighted sum of the pairwise interactions in an atomic cluster (centered around the $i$th atom) projected onto $\Phi_{f_m}$.

Similarly, we can define the general expression of the three-body cluster basis functions for any local clusters of three-body interactions as
\begin{equation}\label{eq:3bcb}
    \begin{aligned}
        &\,\hat{V}_{\rm 3b}[f] \Bigl(\mathcal{N}_i;\,{\bm{\omega}}^{\braket{\mathfrak{C}_3 ( \mathfrak{A} )}},\bm{\theta_{\Phi_m}},\bm{\theta_{\Phi_n}},\left\{\mathcal{A}_{q,ijk}\,|\,j;k\right\} \Bigr) \\
        \equiv\,& \sum_{j<k} \omega^{\braket{e_i, e_j, e_k}}L_{f_m,f_n,ijk}\,\mathcal{A}_{q,ijk}
    \end{aligned}
\end{equation}
where $L_{f_m,f_n,ijk}$ is a symmetrized three-body product based on two localized two-body functions $\Phi_{\!f_m}$ and $\Phi_{\!f_n}$:
\begin{equation}\label{eq:s2b2b}
    \begin{aligned}
        &\, L_{f_m,f_n, ijk} \\
      = &\, L_{f_m,f_n}\bigl(\bm{r_{ij}},\,\bm{r_{ik}};\, \bm{\theta_{\Phi_m}},\,\bm{\theta_{\Phi_n}}\bigr)\\
    \equiv&\, \frac{1}{2} \Bigr(\Phi_{f_m}\! \bigl( \bm{r_{ij}};\bm{\theta_{\Phi_m}}) \, 
              \Phi_{f_n}\!\bigl( \bm{r_{ik}};\bm{\theta_{\Phi_n}}\bigr) \bigl( 2-\delta_{e_je_k} \bigr) \,+\, \\
        &\, \phantom{\frac{1}{2} \Bigr(}\Phi_{f_n}\!\bigl( \bm{r_{ij}};\bm{\theta_{\Phi_n}}\bigr) \, 
              \Phi_{f_m}\!\bigl( \bm{r_{ik}};\bm{\theta_{\Phi_m}}\bigr)\delta_{e_je_k} \Bigr).
    \end{aligned}
\end{equation}
$L_{f_m,f_n,ijk}$ imposes the invariance of three-body interaction under the spatial swapping of two same-element atoms in any three-body composition Y-X-Y, where X represents the center atom.  Additionally, a three-body angular descriptor function
\begin{equation}
    \mathcal{A}_{q,ijk} \!=\! \mathcal{A}_q(\bm{r_{ij}},\,\bm{r_{ik}};\,\bm{A_q}) \!\equiv\! 1 \!+\! \sum_{p=1}^{N_p} A_{p,q}\!\left(\!\frac{\bm{r_{ij}}\cdot\bm{r_{ik}}}{r_{ij}\,r_{ik}}\!\right)^p
\end{equation}
is included in Equation~(\ref{eq:3bcb}).

Finally, we introduce the concept of many-body cluster bases for designing composable interatomic potential models. Formally, a finite many-body cluster basis set is defined as
\begin{equation}
    \mathfrak{F} ( \bm{\theta}_{\mathfrak{F}} ) \equiv
    \bigcup_{n\in\mathfrak{D}} 
    \left\{ \psi_{n{\rm b},\nu} 
    \,|\,\nu\!=\!1,\dots,N_{n{\rm b}}
    \right\}
\end{equation}
with $\mathfrak{D}$ as all the (unique) $n$-body orders for $\psi_{n{\rm b},\nu}$ and $\bm{\theta}_{\mathfrak{F}}$ as the basis set parameters defined by 
\begin{equation}
    \bm{\theta}_{\mathfrak{F}}
    \equiv\hspace{-1em} 
    \bigcup_{n\in\mathfrak{D}\!,\, 1 \leq\nu\leq N_{n{\rm b}}} 
    \hspace{-1em}\bm{\theta_{\psi_{n{\rm b},\nu}}}
\end{equation}
where $\bm{\theta_{\psi_{n{\rm b},\nu}}}$ is the set of parameters characterizing $\psi_{n{\rm b},\nu}$, in which each element is distinguished by its associated symbol (including subscript and superscript) to be properly assigned a new value. For a given $n$, $N_{n {\rm b}}$ determines the number of $n$-body cluster basis functions in $\mathfrak{F}$. Consequently, the cardinality of $\mathfrak{F}$, i.e., the basis set size is
\begin{equation}
    N_{\mathfrak{F}} \equiv \vert \mathfrak{F} \vert = \sum_{n\geq2} N_{n{\rm b}}.
\end{equation}

Since the basis set parameters are the union of all the basis function parameters, different basis functions within the same basis set may (partially) share parameters. In other words, parameter symbols may be reused (thus indistinguishable) across multiple basis functions. Particularly, for Gaussian-type many-body cluster basis sets $\mathfrak{F}_{{\rm 2b},\mathcal{G}}$, we reuse primitive parameters to impose correlations among the exponent coefficients by adopting an even-tempered approach~\cite{feller1979systematic, cherkes2009spanning} from quantum chemistry basis set design. We first define a composite parameter $\bm{\gamma}$ from a pair of positive real numbers:
\begin{equation}\label{eq:et}
    \bm{\gamma} = \{\alpha_0,\,\beta_0\} \quad \text{with} \quad \beta_{\nu} \coloneqq \tilde{\beta}_{\nu}(\bm{\gamma}) \equiv \alpha_0{\beta_0}^{\nu-1}.
\end{equation}
Then, we define an even-tempered Gaussian-type two-body cluster basis function as 
\begin{equation}\label{eq:eg2b}
    \psi_{2{\rm b},\nu} \bigl( \cdot ;\, \bm{\theta_{\psi_{2{\rm b},\nu}}}\bigr) \coloneqq V_{\rm 2b}[\mathcal{G}] \left(\cdot ;\, {\bm{\omega}}^{\braket{\mathfrak{C}_2(\mathfrak{A})}},\,\tilde{\beta}_{\nu}(\bm{\gamma}) \right)
\end{equation}
with 
\begin{equation}
    \bm{\theta_{\psi_{2{\rm b},\nu}}} \coloneqq {\bm{\omega}}^{\braket{\mathfrak{C}_2(\mathfrak{A})}} \cup \bm{\gamma} .
\end{equation}
For systems with a single element type X, where $\mathfrak{C}_2(\mathfrak{A})$ only contains one element $\rm \braket{X,X}$, we can further reduce the size of $\bm{\theta_{\psi_{2{\rm b},\nu}}}$ by simplifying Equation~(\ref{eq:eg2b}) into 
\begin{equation}\label{eq:gtmbc}
    \psi_{2{\rm b},\nu} \bigl( \cdot ;\, \bm{\gamma}\bigr) \coloneqq V_{\rm 2b}[\mathcal{G}] \Bigl(\cdot ;\, \{1\},\,\tilde{\beta}_{\nu}(\bm{\gamma}) \Bigr)
\end{equation}
by setting ${\bm{\omega}}^{\braket{\mathfrak{C}_2(\mathfrak{A})}}$ to $\{1\}$.

\section{The exponentiated pair-cluster model and Kolmogorov--Arnold representations}\label{app:kar}
Kolmogorov--Arnold representation (KAR) theorem~\cite{kolmogorov1961representation, braun2009constructive} states that there always exists a representation for an arbitrary continuous function $f:\,[0,\,1]^n \to \mathbb{R}$ in terms of a finite number of univariate functions:
\begin{equation}\label{eq:kar}
    f(\bm{x}) = \sum_{q=1}^{2n+1}
    g_q\bigl(\sum_{p=1}^n h_{pq}(x_p)\bigr).
\end{equation}
Due to the generality of the KAR theorem, it has been considered a potential approach for designing universal function approximators~\cite{lin1993realization, koppen2002training, lai2021kolmogorov}. However, one challenge the KAR theorem poses is that it does not provide the explicit solution of the proper inner functions $h_{pq}$ and outer functions $g_q$ for the representation. Past literature~\cite{girosi1989representation, poggio2020theoretical} has pointed out that the solved univariate functions are often unsmooth and do not hold simple analytic forms, which hinders the practicality of the KAR theorem for numerical applications.

One way to circumvent the potential instability of the univariate functions in Equation~(\ref{eq:kar}) is by lifting the constraint on the outer-layer dimension ($2n+1$) spanned by $g_q$ in exchange for a pre-selected set of smooth functions as the inner and outer functions. Formally, we define a Kolmogorov--Arnold operator $\hat{\mathcal{K}}$ that generates a bounded continuous multivariate scalar function $f:\,\mathbb{R}^{N_{\rm in}} \to \mathbb{R}$ as:
\begin{equation}\label{eq:gkar}
    \begin{aligned}
        &f(\bm{x}) \coloneqq \hat{\mathcal{K}}[\mathfrak{B}](\bm{x}) \equiv \sum_q g_q\left(\sum_p h_{pq}(x_p)\right)\\
        &\mathrm{with} \hspace{0.5em} \mathfrak{B} \!=\! \{(h_{pq},\,g_q) \,|\, p\!=\!1, \dots, N_{\rm in}; \,q\!=\!1, \dots, N_{\rm out} \}.\\[8pt]
    \end{aligned}
\end{equation}
Thus, we can reformulate the exponentiated pair-cluster model introduced in Sec.~\ref{sec:nlp} as:
\begin{equation}\label{eq:e2b_2}
    \hat{S}_{{\rm e}2}[\mathfrak{F}_{\rm 2b}] \bigl( 
    \mathcal{N}_i\bigr) \coloneqq\,
    \hat{\mathcal{K}}[\{(h_{\nu\mu},\,g_{\mu}) \,|\, \nu, \,\mu\}]\bigl(\, \overrightarrow{\psi_{2{\rm b}}} (i) \,\bigr)
\end{equation}
with
\begin{equation}\label{eq:e2b_2_2}
    \begin{aligned}
    g_{\mu}(y_{\mu}) =&\, a_{\mu}y_{\mu},\\[4pt]
    h_{\nu\mu}(x_{\nu}) =&\, b_{\nu}\exp(\zeta_\mu x_{\nu}),\\[4pt]
    \overrightarrow{\psi_{2{\rm b}}} (i) =&\, \left[\psi_{2{\rm b},1}\!\bigl( \mathcal{N}_i; \bm{\theta_{\psi_{2{\rm b},1}}} \!\bigr),\!
    {\hbox to 0.8em{.\hss.\hss.}}
    ,\!\psi_{2{\rm b},N_{\rm 2b}}\!\left( \mathcal{N}_i;\bm{\theta_{\psi_{2{\rm b},N_{\rm 2b}}}} \right) \right].\\[8pt]
    \end{aligned}
\end{equation}
Comparing Equation~(\ref{eq:e2b_2}) with the original expression of the exponentiated pair-cluster interaction models, Equation~(\ref{eq:e2b}), one can see that two expressions become interchangeable when values of $c_{\mu\nu}$ form a rank-one coefficient matrix $\bm{C}$ (i.e., $C_{ij} = c_{ij}$) obeying the following identity:
\begin{equation}
    c_{\mu\nu} = a_{\mu}b_{\nu}.
\end{equation}
In general, the $\bm{C}$ matrix in the exponentiated pair-cluster interaction models may have a rank larger than one. Hence, it provides a more flexible (generalized) parameterization of the basis functions compared to the conventional KAR theorem.

\clearpage
\onecolumngrid

\section{Complete training results}\label{app:data}
\begin{table}[ht]
    \setlength{\tabcolsep}{1.8pt}
    \renewcommand{\arraystretch}{1.25}
    \centering
    \begin{tabular}{lrrrrrrr}
        \toprule[1.5pt]
        \multicolumn{1}{l}{\multirow{2}{*}{Configuration}} & \multicolumn{2}{c}{Size} & \multicolumn{1}{c}{\multirow{2}{*}{$\lg\kappa$}} & \multicolumn{4}{c}{RMSE} \\ \cline{2-3}\cline{5-8}
        \multicolumn{1}{c}{} & \multicolumn{1}{c}{$\bm{c}$} & \multicolumn{1}{c}{$\bm{d}$} & \multicolumn{1}{c}{} & \multicolumn{1}{c}{$\measuredangle \bm{F}$} & \multicolumn{1}{c}{$\lVert\bm{F}\rVert$} & \multicolumn{1}{c}{$\bm{F}$} & \multicolumn{1}{c}{$\bar{E}$} \\ \midrule[1.25pt]
        G4  & 5  & 2  & 20.29 & 1.6178 & 0.2592 & 1.6329 & 0.1906  \\
        G6  & 7  & 2  & 22.64 & 1.5319 & 0.4210 & 1.1914 & 0.1670  \\
        G8  & 9  & 2  & 28.38 & 1.4724 & 0.6735 & 0.9879 & 0.1085  \\
        G10 & 11 & 2  & 14.59 & 1.6029 & 0.3991 & 0.6907 & 0.0879  \\
        G12 & 13 & 2  & 23.77 & 1.6203 & 0.4087 & 0.5617 & 0.0707  \\
        G14 & 15 & 2  & 20.81 & 1.6228 & 0.3720 & 0.4969 & 0.0616  \\
        G16 & 17 & 2  & 24.80 & 1.6363 & 0.3742 & 0.4872 & 0.0603  \\
        G18 & 19 & 2  & 29.03 & 1.6641 & 0.3670 & 0.4793 & 0.0596  \\\midrule[1.0pt]
        C8  & 9  & 0  & 12.31 & 1.6521 & 0.3871 & 2.1405 & 0.3019  \\
        C16 & 17 & 0  & 21.82 & 1.6243 & 0.4081 & 0.9153 & 0.0992  \\
        C24 & 25 & 0  & 31.24 & 1.8317 & 0.4531 & 0.6979 & 0.0833  \\
        C32 & 33 & 0  & 31.12 & 1.8888 & 0.4595 & 0.6671 & 0.0779  \\
        C40 & 41 & 0  & 31.66 & 1.9651 & 0.4790 & 0.6525 & 0.0723  \\\midrule[1.0pt]
        E[{\small G4L4}]          & 17 & 6  & 27.77 & 1.5642 & 0.4149 & 0.4444 & 0.0347  \\
        E[{\small G4L6}]          & 25 & 8  & 29.27 & 1.5669 & 0.3572 & 0.3690 & 0.0303  \\
        E[{\small G4L8}]          & 33 & 10 & 31.64 & 1.5626 & 0.3642 & 0.3657 & 0.0297  \\
        E[{\small G6L4}]          & 25 & 6  & 28.73 & 1.5712 & 0.3235 & 0.3009 & 0.0219  \\
        E[{\small G6L5}]          & 31 & 7  & 30.80 & 1.5678 & 0.3220 & 0.2912 & 0.0212  \\
        E[{\small G8L4}]          & 33 & 6  & 25.78 & 1.4841 & 0.3036 & 0.2615 & 0.0170  \\
        E[{\small G8L5}]          & 41 & 7  & 28.72 & 1.5328 & 0.2944 & 0.2482 & 0.0167  \\\midrule[1.0pt]
        N[{\small E[{\fs G6L5}]}] & 31 & 7  & 28.93 & 1.6242 & 0.4675 & 0.3254 & 0.0214  \\
        N[{\small E[{\fs G8L4}]}] & 33 & 6  & 24.82 & 1.5940 & 0.4662 & 0.3154 & 0.0185  \\\bottomrule[1.5pt]
    \end{tabular}
    \caption{The complete training results of single-term models based on different basis sets and architectures. G$\nu$ and C$\nu$ represent $\hat{S}_{\,\rm l2}$ with $\nu$ Gaussian-type 
 two-body cluster basis functions and $\nu$ Chebyshev-type two-body cluster basis functions, respectively. E[G$\nu$L$\mu$] represents the $\hat{S}_{\rm e2}$ with $\nu$ Gaussian-type 
 two-body cluster basis functions and $\mu$ latent space layers. Same notation convention is used for $\hat{S}_{\rm ne2}$ configurations N[E[G$\nu$L$\mu$]]. The error metrics of each model configuration are RMSEs of four aspects: the energy $\bar{E}$ ({\evpatm}), the force $\bm{F}$ ({\evpang}), the force amplitude $\lVert\bm{F}\rVert$ ({\evpang}), and the force angle $\measuredangle \bm{F}$ (unitless), respectively. $\bm{c}$ represents the number of linear coefficients, and $\bm{d}$ represents the number of nonlinear coefficients. $\lg \kappa$ represents the logarithmic condition number of the linear-coefficient Fisher information matrix ($\mathcal{I}$).}
    \label{table:singleTerm}
\end{table}

\begin{table*}[htp]
    \setlength{\tabcolsep}{1.8pt}
    \renewcommand{\arraystretch}{1.25}
    \centering
    \begin{tabular}{lrrrrrrrrr}
        \toprule[1.5pt]
        \multicolumn{1}{l}{\multirow{2}{*}{Configuration}} & \multicolumn{2}{c}{Size} & \multicolumn{3}{c}{$\lg\kappa$}  & \multicolumn{4}{c}{RMSE}\\ \cline{2-6}\cline{7-10}
        \multicolumn{1}{c}{} & \multicolumn{1}{c}{$\bm{c}$} & \multicolumn{1}{c}{$\bm{d}$} & \multicolumn{1}{c}{Total} & \multicolumn{1}{c}{Left} & \multicolumn{1}{c}{Right} & \multicolumn{1}{c}{$\measuredangle \bm{F}$} & \multicolumn{1}{c}{$\lVert\bm{F}\rVert$} & \multicolumn{1}{c}{$\bm{F}$} & \multicolumn{1}{c}{$\bar{E}$} \\ \midrule[1.25pt]
        G4  $\times$ G4                 & 11 & 4  & 29.64 & 28.34 & 14.64 & 1.5732 & 0.3085 & 1.3246 & 0.1690 \\
        G6  $\times$ G4                 & 13 & 4  & 29.94 & 29.29 &  7.04 & 1.6575 & 0.4420 & 0.7841 & 0.0700 \\
        G6  $\times$ G6                 & 15 & 4  & 27.22 & 24.88 & 14.73 & 1.6776 & 0.4529 & 0.5649 & 0.0473 \\
        G8  $\times$ G6                 & 17 & 4  & 23.48 & 21.33 & 15.03 & 1.5567 & 0.3461 & 0.3292 & 0.0280 \\
        G8  $\times$ G8                 & 19 & 4  & 31.34 & 16.08 & 28.35 & 1.5677 & 0.3290 & 0.3159 & 0.0282 \\
        G10 $\times$ G10                & 23 & 4  & 30.07 & 24.76 & 27.83 & 1.5871 & 0.3331 & 0.3094 & 0.0267 \\\midrule[1.0pt]
        G8  $\times$ E[{\small G4L4}]   & 27 & 8  & 29.82 & 25.98 & 17.25 & 1.4902 & 0.3162 & 0.2581 & 0.0173 \\
        G8  $\times$ E[{\small G4L6}]   & 35 & 10 & 28.45 & 24.79 & 28.01 & 1.4667 & 0.3110 & 0.2440 & 0.0150 \\
        G8  $\times$ E[{\small G4L8}]   & 43 & 12 & 32.89 & 24.38 & 31.51 & 1.4802 & 0.3067 & 0.2386 & 0.0149 \\
        G8  $\times$ E[{\small G6L4}]   & 35 & 8  & 28.50 & 26.43 & 27.64 & 1.5344 & 0.2500 & 0.2098 & 0.0155 \\
        G8  $\times$ E[{\small G6L5}]   & 41 & 9  & 29.87 & 24.38 & 28.83 & 1.5403 & 0.2501 & 0.2069 & 0.0151 \\
        G8  $\times$ E[{\small G8L4}]   & 43 & 8  & 31.15 & 27.76 & 28.78 & 1.5173 & 0.2420 & 0.2033 & 0.0148 \\
        G8  $\times$ E[{\small G8L5}]   & 51 & 9  & 30.52 & 27.47 & 28.45 & 1.4829 & 0.2348 & 0.1940 & 0.0147 \\
        G10 $\times$ E[{\small G6L5}]   & 43 & 9  & 30.58 & 14.58 & 30.82 & 1.5205 & 0.2519 & 0.2066 & 0.0162 \\
        G10 $\times$ E[{\small G8L4}]   & 45 & 8  & 28.30 & 14.60 & 27.28 & 1.5040 & 0.2453 & 0.1870 & 0.0132 \\
        G12 $\times$ E[{\small G8L4}]   & 47 & 8  & 25.86 & 21.41 & 23.85 & 1.4658 & 0.2527 & 0.2011 & 0.0165 \\\midrule[1.0pt]
        E[{\small G6L5}] $\!\!+\!\!$ N[{\small E[{\fs G6L5}]}] & 61 & 12 & 28.16 & 25.32 & 26.57 & 1.4729 & 0.2066 & 0.1813 & 0.0128 \\
        E[{\small G8L4}] $\!\!+\!\!$ N[{\small E[{\fs G8L4}]}] & 65 & 10 & 23.97 & 21.44 & 22.87 & 1.4597 & 0.1975 & 0.1718 & 0.0126 \\\bottomrule[1.5pt]
    \end{tabular}
    \caption{The complete training results of composable model configurations based on dual-term architectures. With $\nu$ denoting the number of Gaussian-type two-body cluster basis functions and $\mu$ denoting the number of latent space layers, G$\nu_{\rm L}$$\times$G$\nu_{\rm R}$ represent the model configurations based on $\hat{P}_{\, {\rm l2l2}}$; G$\nu_{\rm L}$$\times$E[G$\nu_{\rm R}$L$\mu$] represent the model configurations based on $\hat{P}_{\, {\rm l2e2}}$; E[G$\nu$L$\mu$]$+$N[E[G$\nu$L$\mu$]] represent the model configurations based on $\hat{P}_{\rm ene2}$. The error metrics of each model configuration are RMSEs of four aspects: the energy $\bar{E}$ ({\evpatm}), the force $\bm{F}$ ({\evpang}), the force amplitude $\lVert\bm{F}\rVert$ ({\evpang}), and the force angle $\measuredangle \bm{F}$ (unitless), respectively. $\bm{c}$ represents the number of linear coefficients, and $\bm{d}$ represents the number of nonlinear coefficients. $\lg \kappa$ represents the logarithmic condition number of the linear-coefficient Fisher information matrix ($\mathcal{I}$). The local logarithmic $\mathcal{I}$ condition numbers of left-term and right-term submodels inside each dual-term potential model are also included.}
    \label{table:dualTerm}
\end{table*}

\end{document}